\documentclass[12pt,a4paper]{article} 
\usepackage[english]{babel} 
\usepackage[utf8]{inputenc} 
\usepackage[breaklinks=true]{hyperref}
\usepackage{graphicx}
\usepackage{amssymb}
\usepackage{amsthm}
\usepackage{amsmath}
\usepackage{microtype}
\usepackage{scalefnt}
\usepackage{booktabs} 
\usepackage{multicol}         
\usepackage{here}
\usepackage{mathtools}
\usepackage{url}
\usepackage{xurl}
\usepackage[bottom=2cm, left=3cm, right=2cm, top=3cm]{geometry} %
\usepackage{multirow} 
\usepackage{breqn}
\usepackage{bm} 
\usepackage[]{natbib} 
\usepackage{float} 
\usepackage{graphicx} 
\theoremstyle{definition}

\title{The modified odd Burr XII-G family of distributions: Properties and applications}
\author{Alexsandro A. Ferreira \\
Universidade Federal de Pernambuco, Brazil. 
\\ alexsandro.ferreira.aaf@gmail.com \\
Gauss M. Cordeiro \\
Universidade Federal de Pernambuco, Brazil. 
\\ gauss@de.ufpe.br\\}
\date{}

\begin{document}
\maketitle

\begin{abstract}
\noindent The modified odd Burr XII-G family is developed, capable of incorporating bimodal and bathtub shapes in its baseline distributions, with properties derived from the exponentiated-G class. A regression model is developed within this family. The parameters are estimated by maximum likelihood, and simulations are performed to verify their consistency. The usefulness of the proposals is demonstrated by means of three real data sets.\\

\end{abstract}

\section{Introduction}
	The flexibility of classical distributions, such as Weibull, gamma, and exponential, has been a significant focus of research in recent decades. Several studies have introduced new parameters into these distributions to improve their modeling capabilities and adapt them to various types of data. A notable example is the approach proposed by \citet{mudholkar1993}, which adds an extra parameter to the Weibull distribution, allowing it to handle a bathtub-shaped failure rate function (hrf). In this context, several other studies have made substantial contributions, including those by \citet{MOE1997}, \citet{Gupta1998}, \citet{Eugene2002}, \citet{ZOGRAFOS2009}, \citet{Cordeiro2011}, \citet{Alexander2012}, \citet{Cordeiro2013d}, \citet{Alzaatreh2013c}, \citet{Bourguignon2014}, \citet{alizadeh2015new}, \citet{chipepa2019}, \citet{baharith202}, and \citet{tlhaloganyang2022gamma}, among many others. 
	
Consider $G(x) = G(x;\bm{\xi})$ as the cumulative distribution function (cdf) of any given baseline distribution, where $\bm{\xi}$ is the parameter vector for $G(x)$. Let $r(v)= r(v;\bm{\psi})$ denote the probability density function (pdf) of a random variable $V \in [\,c,d\,]$ (for $-\infty \leq c < d \leq \infty$), having parameter vector $\bm{\psi}$. Next, take a function $W[G(x)] \in [\,c,d\,]$ of $G(x)$. This function is assumed to be differentiable, monotonically non-decreasing, and $W[G(x)] \rightarrow c$ as $x \rightarrow -\infty$ and $W[G(x)] \rightarrow d$ as $x \rightarrow \infty$. Under these conditions, the transformed-transformer (T-X) family has a cdf in the form \citep{Alzaatreh2013c}.
\begin{align}\label{txcdfMOB}
F(x) = F(x;\bm{\xi}, \bm{\psi}) = \int^{W[G(x;\bm{\xi})]}_0 r(v;\bm{\psi})\,\text{d}v\,.
\end{align}
	
	Based on (\ref{txcdfMOB}),  the modified odd Burr XII-G (MOBXII-G) family is proposed, which is obtained using $W[G(x)] =\frac{G(x)}{1 - G(x)[1+G(x)]/2}$ proposed by \citet{chesneau2020modified}, and $r(v) = \tau \, \lambda\, v^{\tau - 1} (1 + v^\tau)^{-(\lambda  + 1)}$ having the pdf of the Burr XII distribution. Thus, the cdf of the new family takes the form (for $x \in \text{I\!R}$, and $\tau, \lambda > 0$)
\begin{align}\label{MOBXIIcdf}
F(x) = 1 - \left(1 + \left[\frac{2\,G(x)}{2 - G(x)[1+G(x)]}\right]^\tau\right)^{-\lambda}\,,
\end{align}
and its pdf can be expressed as
\begin{align}\label{MOBXIIpdf}
f(x) = &\frac{2^{\tau}\,\tau\,\lambda\,g(x)\,[2 + G(x)^2]\,G(x)^{\tau-1}}{\{2 - G(x)[1+G(x)]\}^{\tau+1}}\left\{1 + \left[\frac{2\,G(x)}{2 - G(x)[1+G(x)]}\right]^\tau\right\}^{-(\lambda+1)}\,,\,\, x \in \text{I\!R} \, .
\end{align}
	
	The hrf associated with (\ref{MOBXIIpdf}) is
\begin{align}\label{MOBXIIGhrf}
h(x) = \frac{2^{\tau}\,\tau\,\lambda\,g(x)\,[2 + G(x)^2]\,G(x)^{\tau-1}}{\{2 - G(x)[1+G(x)]\}^{\tau+1}}\left\{1 + \left[\frac{2\,G(x)}{2 - G(x)[1+G(x)]}\right]^\tau\right\}^{-1}\,.
\end{align}

	Henceforth, let $X \sim \text{MOBXII-G}(\tau,\lambda,\bm{\xi})$ be a random variable with pdf (\ref{MOBXIIpdf}). Note that for $\lambda = 1$, the MOBXII-G reduces to modified odd log-logistic-G (MOLL-G), with only one extra parameter. 
	
	The introduction of the MOBXII-G family is primarily driven by its enhanced flexibility compared to other established families, particularly the Kumaraswamy-G (K-G) \citep{Cordeiro2011} and beta-G (B-G) \citep{Eugene2002}. These two families have been extensively studied in the literature, giving rise to over 100 distinct distributions each, as observed by \citet{selim2020distributions}. Their widespread application and acceptance underscore their robustness and adaptability in fitting diverse data sets.
	
	However, the MOBXII-G family stands out as a particularly attractive alternative in this area due to its superior flexibility. This is evident in its ability to handle bimodal and bathtub shapes in its baseline models, as illustrated in Figures \ref{FMOBXIIW}, \ref{FMOBXIIK}, and \ref{FMOBXIIN}, which show several examples of these shapes. As a result, the MOBXII-G family can more efficiently model real-world data with these characteristics, as proven by the empirical analysis in Section \ref{sec5MOBXII}.
	
	The rest of the article unfolds as follows. Section \ref{sec1MOBXII} discusses three special models of the new family and Section \ref{sec2MOBXII} describes its main properties. Section \ref{sec3MOBXII} provides a regression model for a special case of the new family, and simulations are reported in Section \ref{sec4MOBXII}. Section \ref{sec5MOBXII} presents three applications to real data, and some conclusions are addressed in Section \ref{sec6MOBXII}.

\section{Some MOBXII-G models}\label{sec1MOBXII}
	This section presents the pdfs for three members of the new family. Their cdfs and hrfs are readily obtained using Equations (\ref{MOBXIIcdf}) and (\ref{MOBXIIGhrf}), respectively. The figures presented here are generated in \texttt{R} \citep{R2023}.
	
\subsection{The modified odd Burr XII Weibull (MOBXIIW)}
	Consider the Weibull distribution ($\alpha, \beta > 0$) in (\ref{MOBXIIpdf}) with cdf (for $x > 0$)
\begin{align*}
G(x) = 1 - \text{e}^{- (x/\beta)^\alpha}.
\end{align*}

	Thus, the density of the MOBXIIW becomes
\begin{align}\label{MOBXIIWpdf}
f(x) =\,\,&\frac{ 2^{\tau}\,\tau\,\lambda\,\alpha\,\beta^{-\alpha} x^{\alpha - 1}\text{e}^{-(x/\beta)^\alpha}\left[2 + \left(1 - \text{e}^{-(x/\beta)^\alpha}\right)^2\right]\left(1 - \text{e}^{-(x/\beta)^\alpha}\right)^{\tau-1}}{\left[2 - \left(1 - \text{e}^{-(x/\beta)^\alpha}\right)\left(2 - \text{e}^{-(x/\beta)^\alpha}\right)\right]^{\tau+1}}\nonumber\\
&\times\left\{1 + \left[\frac{2\left(1 - \text{e}^{-(x/\beta)^\alpha}\right)}{2 - \left(1 - \text{e}^{-(x/\beta)^\alpha}\right)\left(2 - \text{e}^{-(x/\beta)^\alpha}\right)}\right]^\tau\right\}^{-(\lambda+1)}
\end{align}

\begin{figure}[!ht]
		\begin{center}
			\begin{minipage}[c]{0.48\linewidth}
				\centering
				\includegraphics[width=\textwidth, height=7cm]{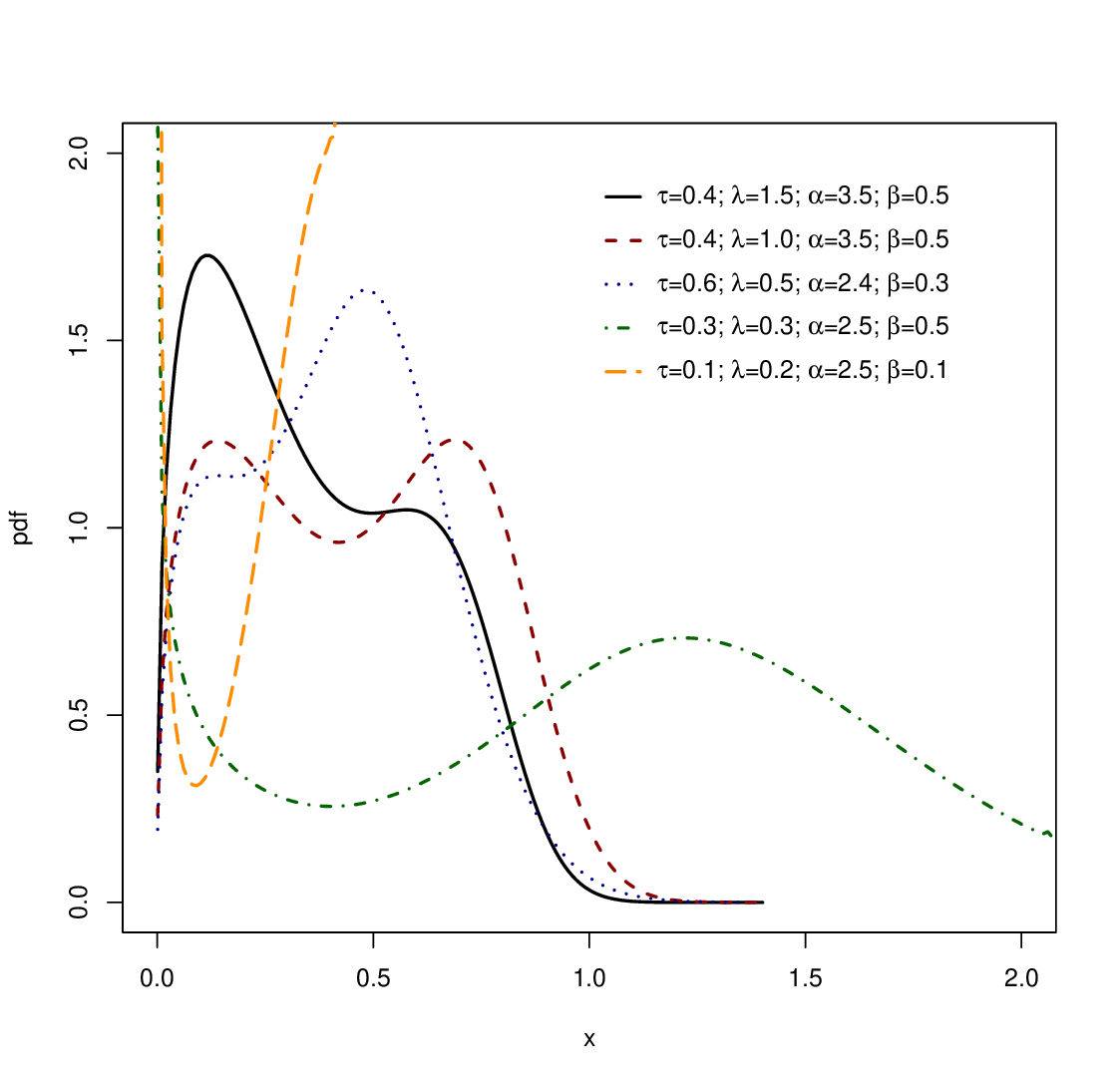}
			\end{minipage}
\hspace{.3cm}
			\begin{minipage}[c]{0.48\linewidth}
				\centering
				\includegraphics[width=\textwidth, height=7cm]{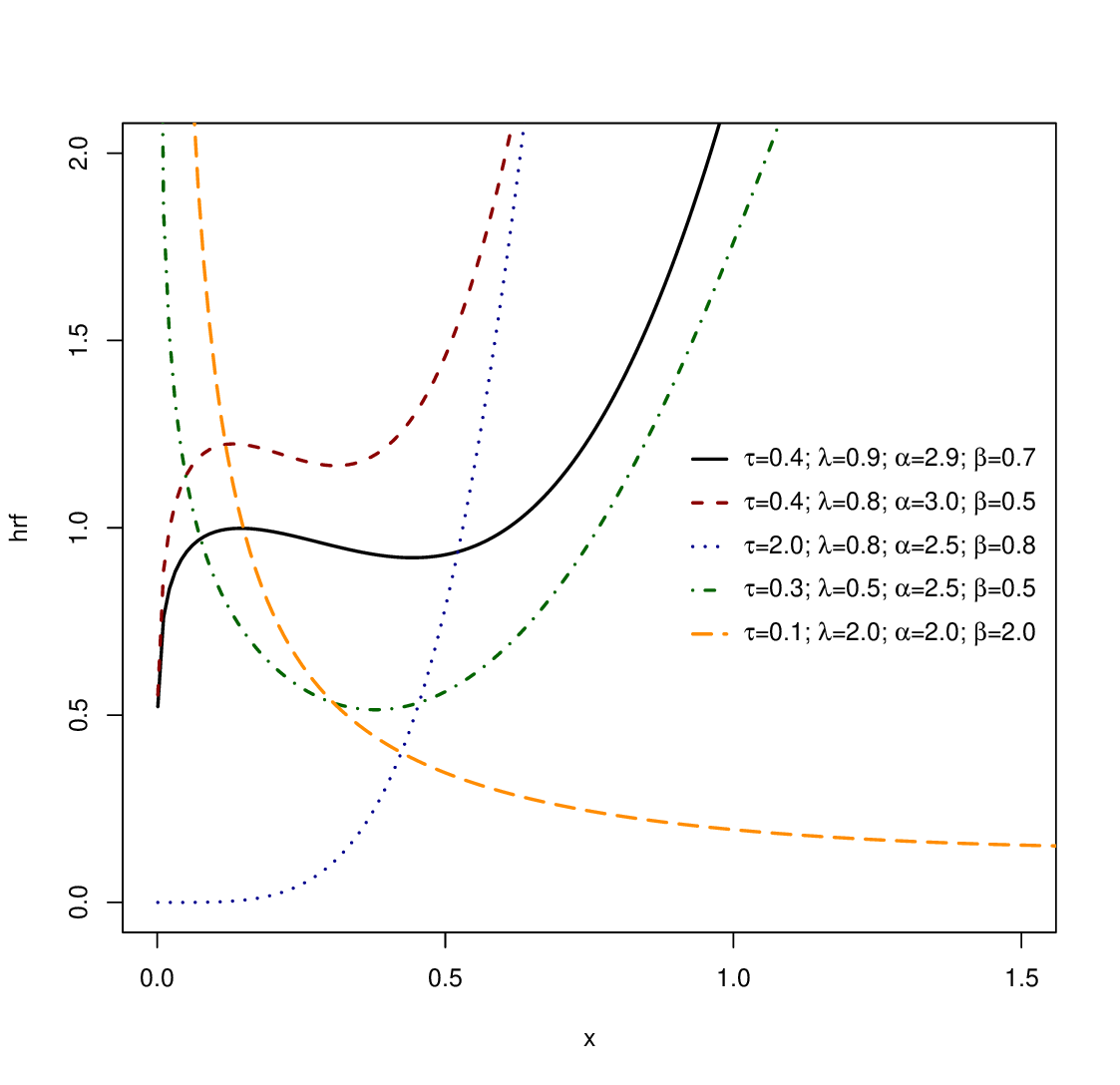}
			\end{minipage}
		\end{center}
		\caption{Density and hrf of $\text{MOBXIIW}(\tau,\lambda,\alpha,\beta)$.}
	\label{FMOBXIIW}
	\end{figure}

\subsection{The modified odd Burr XII Kumaraswamy (MOBXIIK)}
The Kumaraswamy cdf with parameters $a,b>0$ is expressed by (for $0<x<1$) 
\begin{align}\label{NFWK}
G(x) = 1 - (1 - x^a)^b\,.
\end{align}
	
	From (\ref{MOBXIIpdf}) and (\ref{NFWK}), the density of the MOBXIIK can be expressed as
\begin{align*}
f(x) = & \,\,\frac{2^{\tau}\,\tau\,\lambda\,a\,b\,x^{a-1}(1-x^a)^{b-1}\left\{2 + \left[1 - \left(1 - x^a\right)^b\right]^2\right\}\left[1 - \left(1 - x^a\right)^b\right]^{\tau-1}}{\left\{2 - \left[1 - \left(1 - x^a\right)^b\right]\left[2 - \left(1 - x^a\right)^b\right]\right\}^{\tau+1}}\nonumber\\
&\times \left\{1 + \left(\frac{2\left[1 - \left(1 - x^a\right)^b\right]}{2 - \left[1 - \left(1 - x^a\right)^b\right]\left[2 - \left(1 - x^a\right)^b\right]}\right)^\tau\right\}^{-(\lambda+1)}\,.
\end{align*}

\begin{figure}[!ht]
		\begin{center}
			\begin{minipage}[c]{0.48\linewidth}
				\centering
				\includegraphics[width=\textwidth, height=7cm]{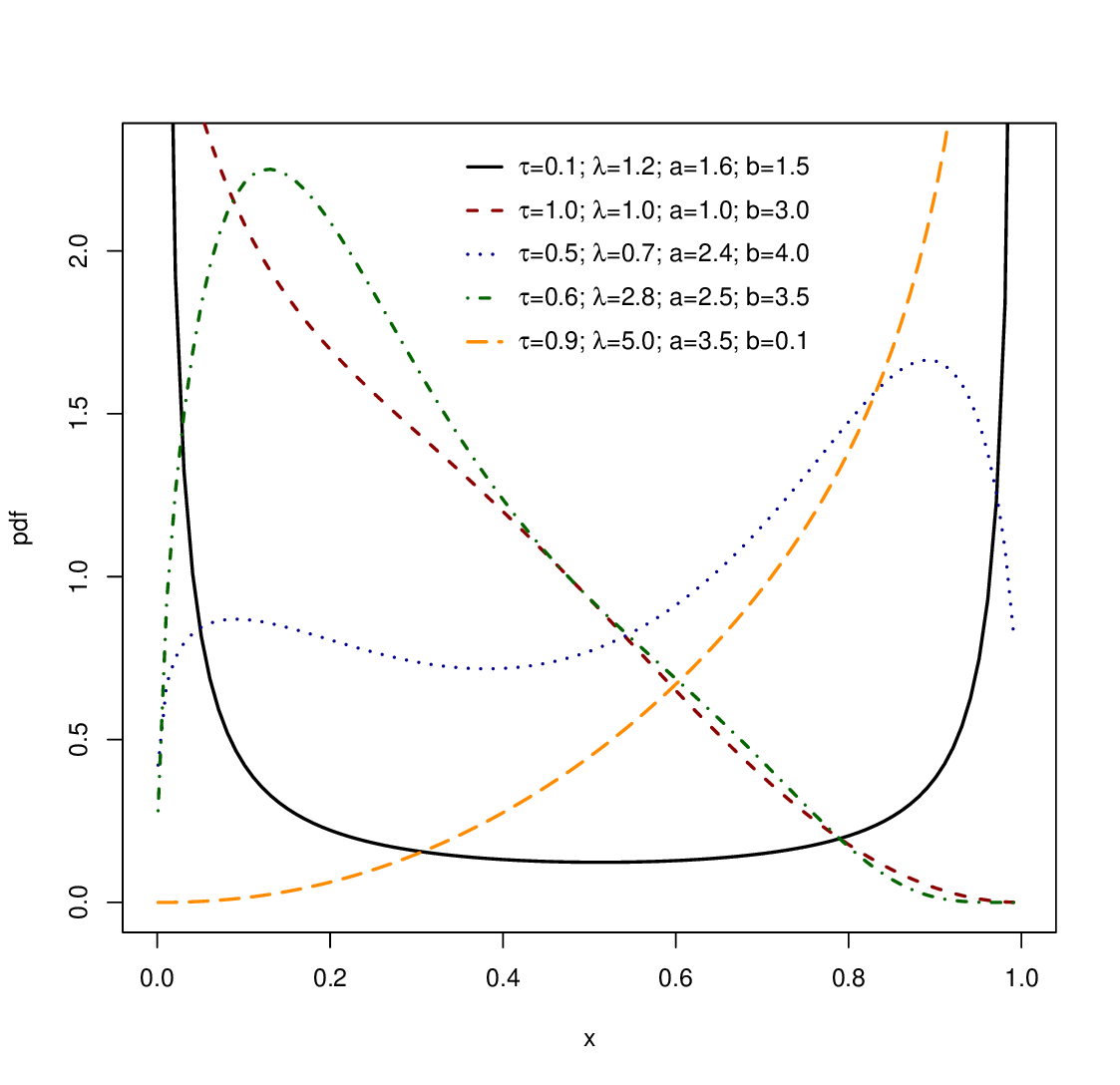}
			\end{minipage}
\hspace{.3cm}
			\begin{minipage}[c]{0.48\linewidth}
				\centering
				\includegraphics[width=\textwidth, height=7cm]{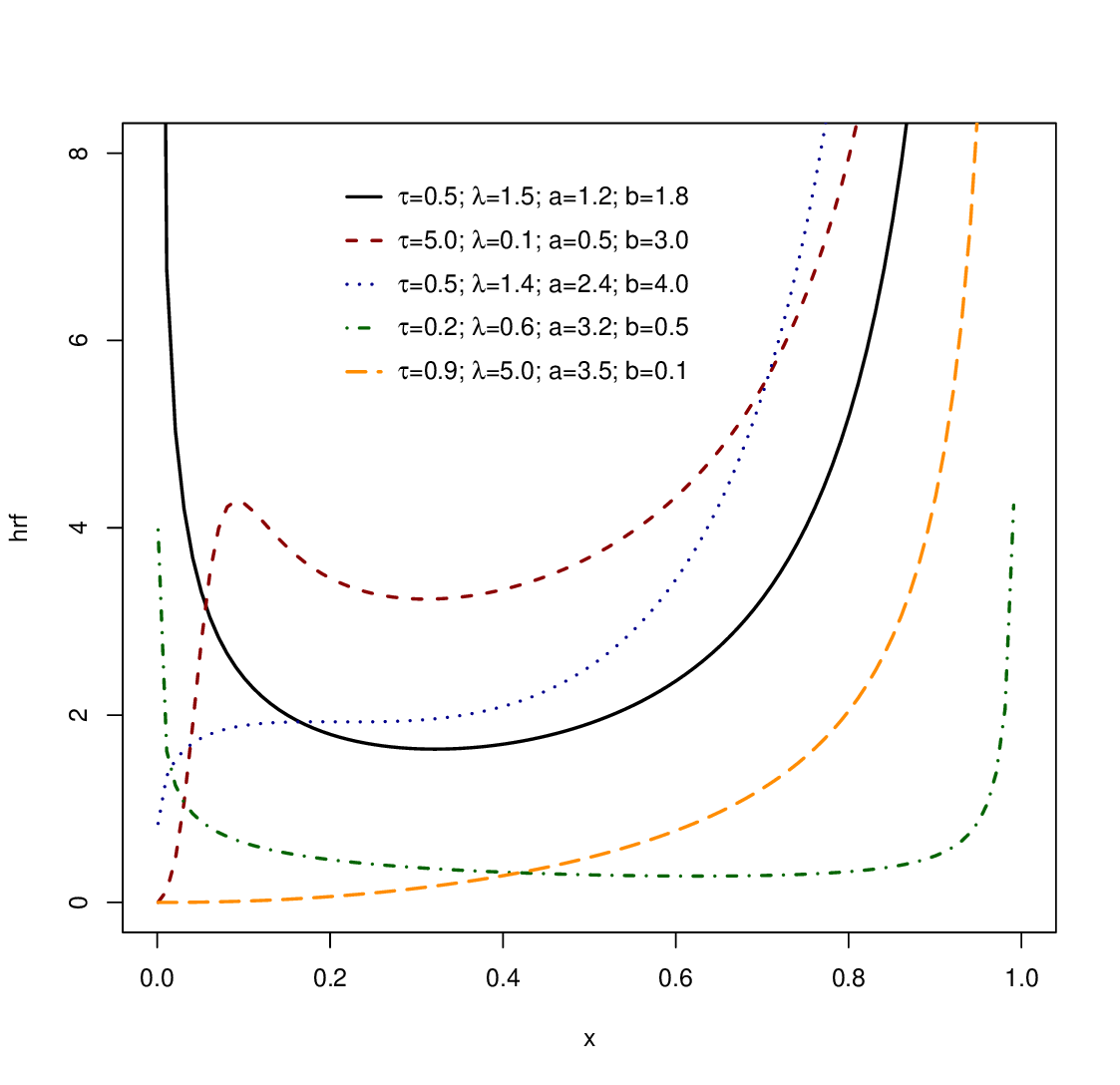}
			\end{minipage}
		\end{center}
		\caption{Density and hrf $\text{MOBXIIK}(\tau,\lambda,a,b)$}
	\label{FMOBXIIK}
	\end{figure}

\subsection{The modified odd Burr XII normal (MOBXIIN)}

	The MOBXIIN density follows from (\ref{MOBXIIpdf}) and the normal baseline $N(\mu, \sigma^2)$ (for $\mu \in \text{I\!R}$, and $\sigma > 0)$ as
\begin{align*}
f(x) = &\frac{2^{\tau}\,\tau\,\lambda\,\phi(z)\,[2 + \Phi(z)^2]\,\Phi(z)^{\tau-1}}{\{2 - \Phi(z)[1+\Phi(z)]\}^{\tau+1}}\left\{1 + \left[\frac{2\,\Phi(z)}{2 - \Phi(z)[1+\Phi(z)]}\right]^\tau\right\}^{-(\lambda+1)}\,,
\end{align*}
where $z = (x-\mu)/\sigma$, and $\phi(\cdot)$ and $\Phi(\cdot)$ are the pdf and cdf of the standard normal, respectively.

\begin{figure}[!ht]
		\caption{Density and hrf of $\text{MOBXIIN}(\tau,\lambda,\mu,\sigma)$.}
		\begin{center}
			\begin{minipage}[c]{0.48\linewidth}
				\centering
				\includegraphics[width=\textwidth, height=7cm]{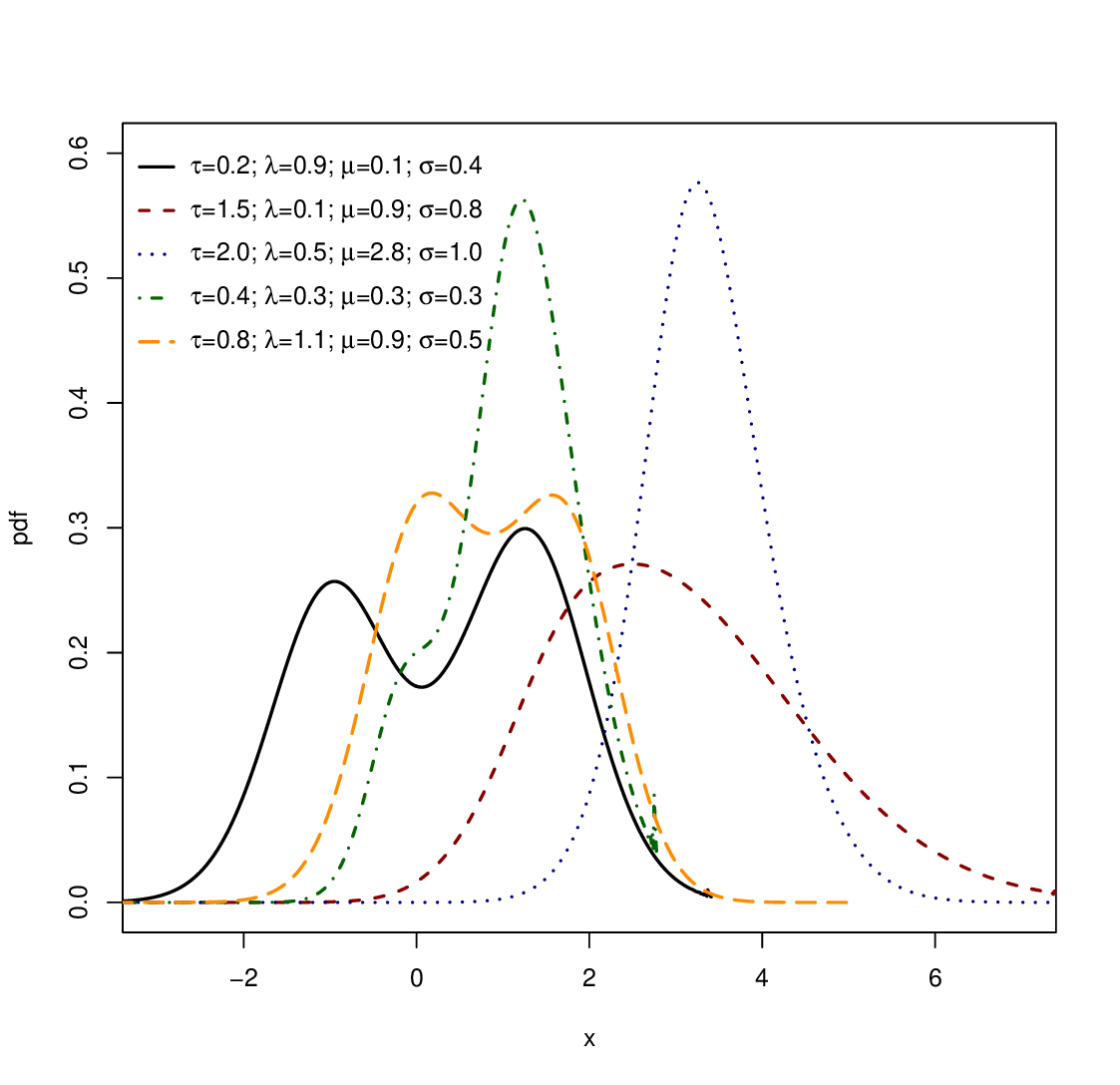}
			\end{minipage}
\hspace{.3cm}
			\begin{minipage}[c]{0.48\linewidth}
				\centering
				\includegraphics[width=\textwidth, height=7cm]{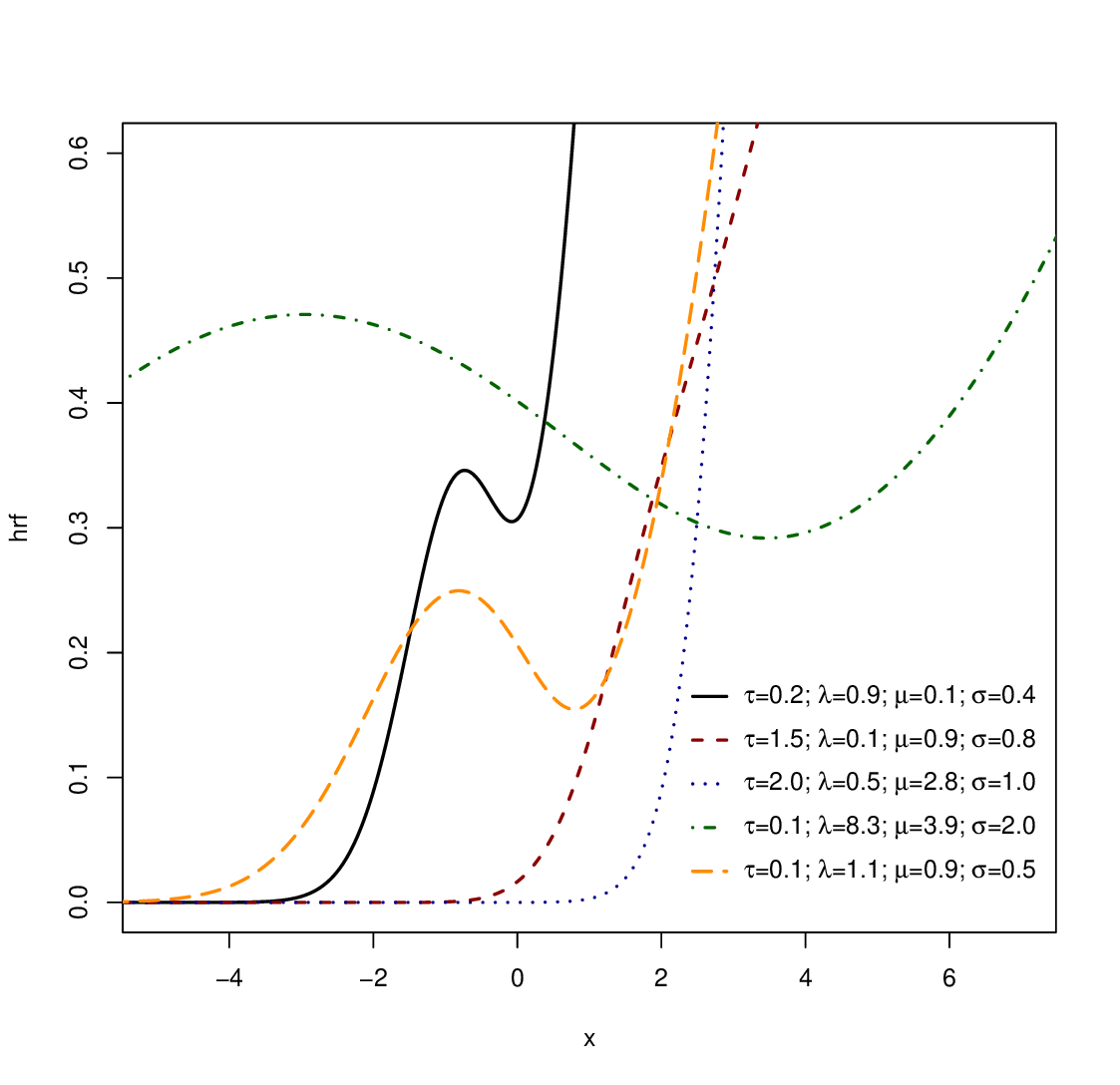}
			\end{minipage}
		\end{center}
		\caption{Density and hrf of $\text{MOBXIIN}(\tau,\lambda,\mu,\sigma)$.}
	\label{FMOBXIIN}
	\end{figure}
		
	The pdfs and hrfs of three new members of the family are illustrated in Figures \ref{FMOBXIIW}, \ref{FMOBXIIK}, and \ref{FMOBXIIN}, respectively. These models can handle a wide range of data, including right-skewed, symmetric, left-skewed, and even bimodal data. Additionally, their hrfs can have various shapes, including unimodal, increasing-decreasing-increasing, or bathtub.
	
\section{Properties}\label{sec2MOBXII}
\subsection{Useful expansions}
	The well-studied exponentiated-G (exp-G) class has a pdf of $\pi_\delta(x) = \delta\,g(x)\,G(x)^{\delta-1}$ (for $\delta > 0$). Several notable distributions fall under this class, including the exp-Weibull \citep{mudholkar1993}, exp-exponential \citep{gupta2001}, exp-Fréchet \citep{Kotz2003}, and exp-gamma \citep{nadarajah2007}, as documented in Table 1 of \citet{Tahir2015}. Thus, the density of the MOBXII-G family admits an expression based on the density of the exp-G class. Initially, the cdf given in (\ref{MOBXIIcdf}) can be expressed as
\begin{align}\label{MOBXIIexp}
F(x) = 1 - \left(\frac{\left\{1 - G(x)[1+G(x)]/2\right\}^\tau}{\left\{1 - G(x)[1+G(x)]/2\right\}^\tau + G(x)^\tau}\right)^\lambda\,.
\end{align}

	Applying the binomial theorem twice, one has
\begin{align}\label{expnumeradorMOB}
\left\{1 - G(x)[1+G(x)]/2\right\}^\tau = \sum^\infty_{m = 0} a_m G(x)^m\,,
\end{align}
where 
\begin{align*}
a_m = a_m(\tau) = \sum_{(i,j)\in I_m} (-1)^{i}2^{-i}\binom{\tau}{i}\binom{i}{j}\,,
\end{align*}
and $I_m = \{(i,j) \in \mathbb{N}^{2}_0\mid i+j = m, j \leq i\}$, $\mathbb{N}_0 = \{0,1,2,\ldots\}$\,. Again, by the binomial theorem, a power series for $G(x)^\tau$ can be found as
\begin{align}\label{expdenominadorMOB}
G(x)^\tau = \{1 - [1-G(x)]\}^\tau = \sum^\infty_{m = 0} b_m G(x)^m\,,
\end{align}
where
\begin{align*}
b_m = b_m(\tau)= \sum^\infty_{\ell = m}(-1)^{\ell + m}\binom{\tau}{\ell}\binom{\ell}{m}\,.
\end{align*}

	By inserting (\ref{expnumeradorMOB}), and (\ref{expdenominadorMOB}) into (\ref{MOBXIIexp}),
\begin{align*}
F(x) = 1 - \left(\frac{\sum^\infty_{m = 0} a_m G(x)^m}{\sum^\infty_{m = 0} d_m G(x)^m}\right)^\lambda\,,
\end{align*}
where $d_m = a_m + b_m$. Then, the quotient of two power series can be represented as
\begin{align}\label{MOLLexpcdf}
F(x) = 1 - \left(\sum^\infty_{m = 0}\omega_m \, G(x)^m \right)^\lambda\,,
\end{align}
where $\omega_0 = a_0/d_0$, and for $m > 0$,
\begin{align*}
\omega_m = \frac{1}{d_0}\left(a_m - \sum^m_{n = 1}d_n\,\omega_{m-n}\right)\,.
\end{align*}
	
	Following the findings of \citet{munir2013} for a power series raised to a non-zero real number, the cdf of the MOBXII family can be expressed in the form 
\begin{align}\label{MOBXIIexpcdf}
F(x) = 1 - \sum^\infty_{m = 0}\vartheta_m \, G(x)^m\,,
\end{align}
where $\vartheta_0 = \omega_0^\lambda$, and for $m >0$, 
\begin{align*}
\vartheta_m = \frac{1}{m\,\omega_0}\left(\sum^{m-1}_{q=0}\left[\lambda\,m - (\lambda+1)q\right]\vartheta_q\,\omega_{m-q}\right)\,.
\end{align*}

	The pdf of the MOBXII-G family follows by differentiating (\ref{MOBXIIexpcdf}) as
\begin{align}\label{MOBXIIexppdf}
f(x) = \sum^\infty_{m = 0}\, \varphi_{m+1}\,\pi_{m+1}(x)\,,
\end{align}
where $\varphi_{m+1} = -\vartheta_{m+1}$ and $\pi_{m+1}(x)$ is the density of the exp-G class with power $(m+1)$. Thus, Equation (\ref{MOBXIIexppdf}) shows that the density of the new family can be expressed as an infinite mixture of exp-G densities, making it easy to derive its properties. Furthermore, setting $\lambda=1$ in Equation (\ref{MOLLexpcdf}) produces the cdf of the MOLL-G family, from which its pdf can be obtained by differentiation.

\subsection{Quantile function}
The MOBXII-G family offers a straightforward analytical expression for its quantile function. Denoting $Q_{G}(x)$ as the quantile function corresponding to $G(x)$, the quantile function for the MOBXII-G family is formulated as (for $0 < u < 1$)
\begin{align}\label{MOBXIIGqf}
Q_X(u) = Q_G\left( -\frac{1}{2} - \left[(1-u)^{-1/\lambda} - 1\right]^{-1/\tau} + \frac{1}{2}\sqrt{\left(1 + 2\left\{\left[(1-u)^{-1/\lambda} - 1\right]^{-1/\tau}\right\}\right)^2 + 8}\right)\,.
\end{align}
	
	Consequently, the MOBXII-G observations for a given $G(x)$ can be derived directly from Equation (\ref{MOBXIIGqf}), along with its median, by setting $u = 1/2$. Additionally, the Bowley skewness \citep{Kenney1962} and Moors kurtosis \citep{Moors1988} for this family are, respectively,
\begin{align*}
\mathcal{B} = \frac{Q_X(3/4) + Q_X(1/4) - 2Q_X(1/2)}{Q_X(3/4) - Q_X(1/4)}\,,
\end{align*}
and
\begin{align*}
\mathcal{M} = \frac{Q_X(7/8) - Q_X(5/8) + Q_X(3/8) - Q_X(1/8)}{Q_X(6/8) - Q_X(2/8)} \, .
\end{align*}
	
\begin{figure}[!ht]
		\begin{center}
			\begin{minipage}[c]{0.48\linewidth}
				\centering
				\includegraphics[width=\textwidth, height=7cm]{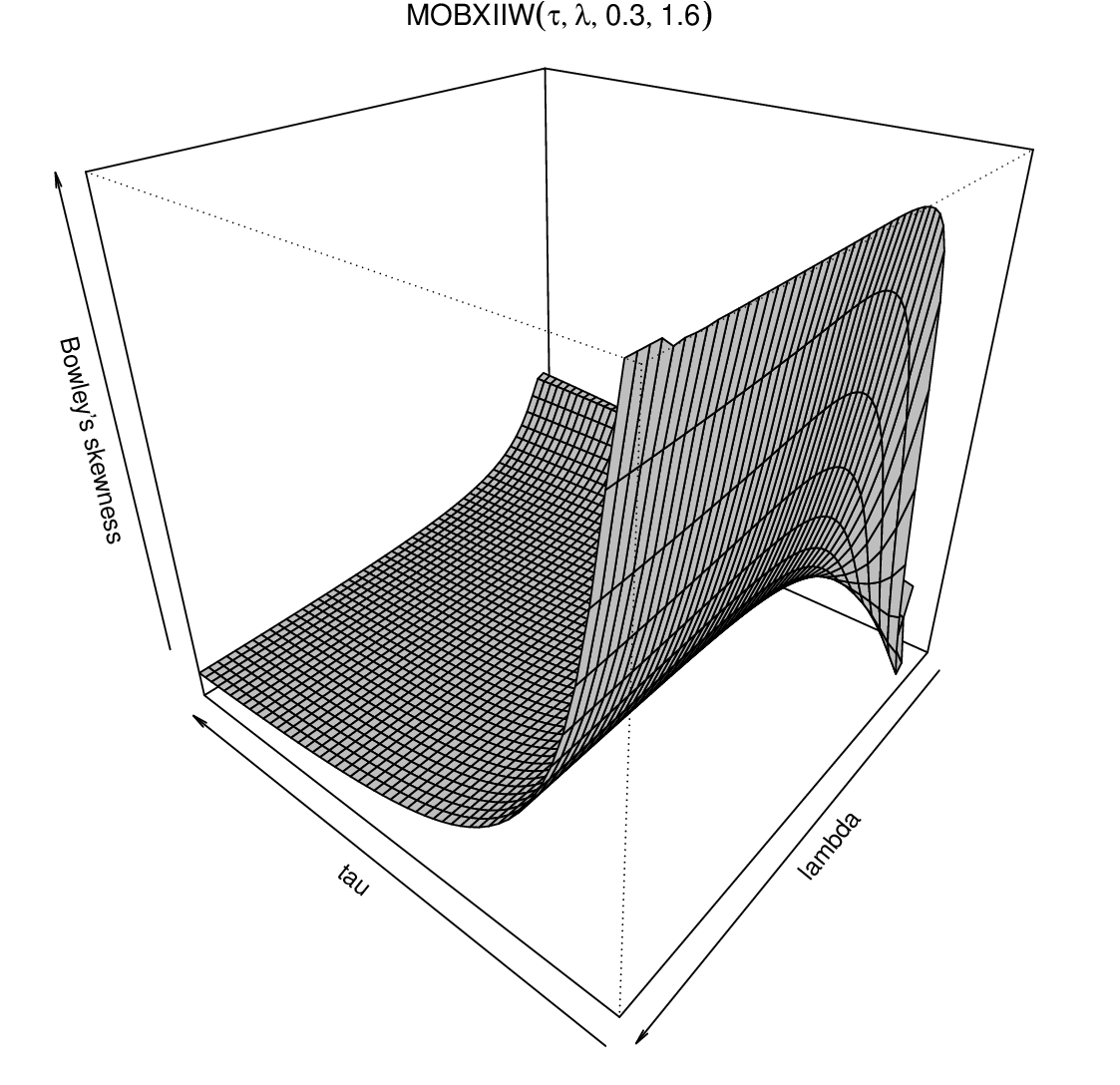}
			\end{minipage}
\hspace{.3cm}
			\begin{minipage}[c]{0.48\linewidth}
				\centering
				\includegraphics[width=\textwidth, height=7cm]{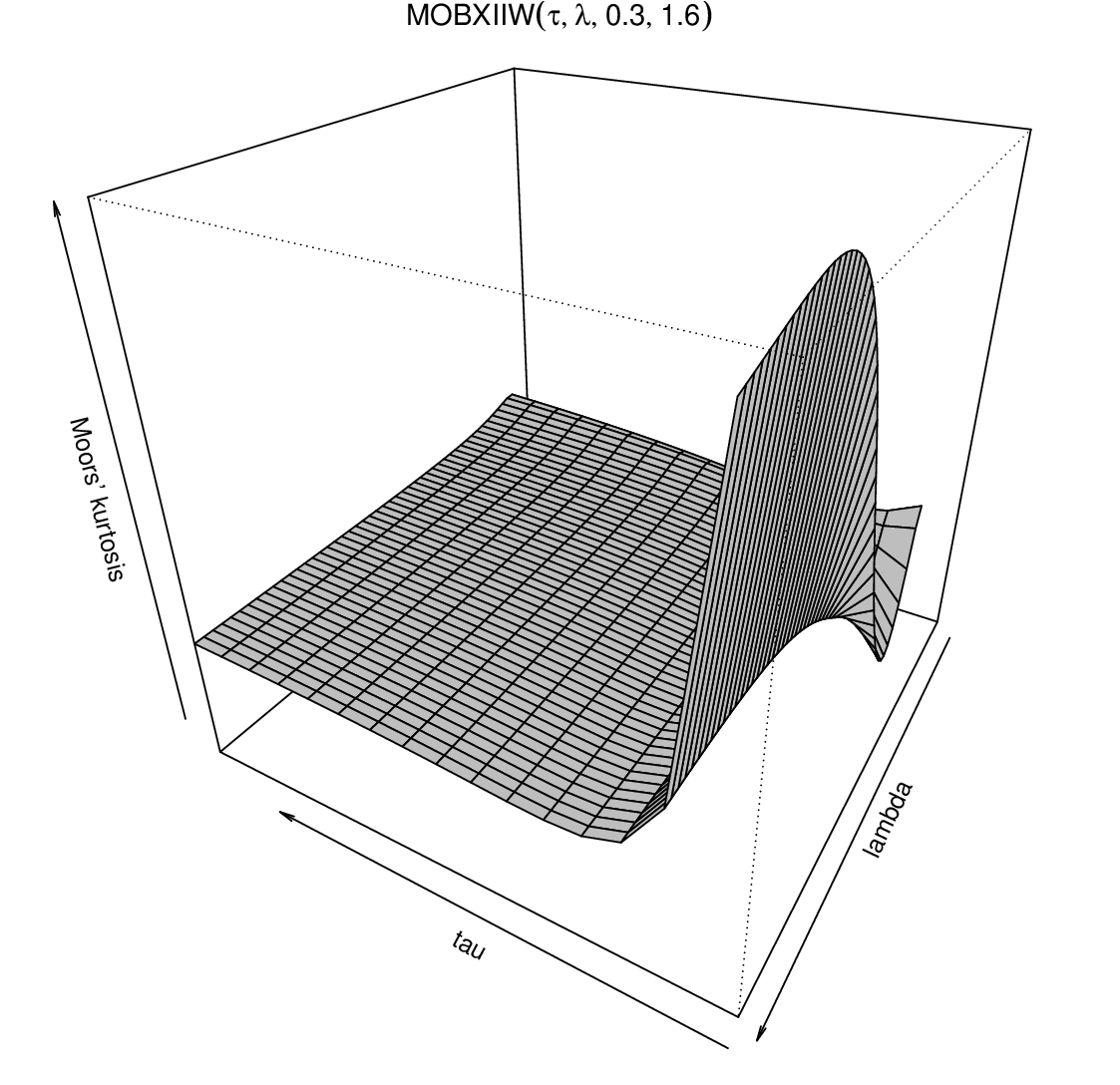}
			\end{minipage}
		\end{center}
		\caption{Skewness and Kurtosis of $\text{MOBXIIW}(\tau,\lambda,\alpha,\beta)$.}
	\label{skewness/kurtosisMOB}
	\end{figure}

	These measures are based on percentiles, making them more resistant to outliers. This offers a significant advantage over moment-based skewness and kurtosis, which are highly sensitive to extreme values. Plots of these measures considering the MOBXIIW distribution (with $\tau$ and $\lambda$ varying) are reported in Figure \ref{skewness/kurtosisMOB}, which indicates that both skewness and kurtosis increase as $\tau$ decreases and $\lambda$ increases.

\subsection{Moments}
The $r$th moment of $X$ can be derived from (\ref{MOBXIIexppdf}) in the form
\begin{align*}
\mu_r^\prime = \text{E}(X^r) = \sum^{\infty}_{m=0} \varphi_{m+1} \, \text{I\!E}(Y^r_{m+1}) = \sum^{\infty}_{m=0} (m+1)\,\varphi_{m+1} \int^1_0 Q_G(u)^r u^{m} \text{d}u\,,
\end{align*}
where $Y_{m+1}$ is the density of the random variable exp-G$(m+1)$.
	
	The $r$th incomplete moment of $X$, $m_r(z) = \int^z_{-\infty}x^r f(x)\text{d}x$, follows from (\ref{MOBXIIexppdf}) as
\begin{align*}
m_r(z) =  \sum^{\infty}_{m=0} \varphi_{m+1} \int^z_{-\infty} x^r \pi_{m+1}(x) \, \text{d}x =  \sum^{\infty}_{m=0} (m+1)\,\varphi_{m+1}\,\int^{G(z)}_0Q_G(u)^r u^{m}\,\text{d}u\,.
\end{align*}
	
\begin{figure}[!ht]
		\begin{center}
			\begin{minipage}[c]{0.48\linewidth}
				\centering
				\includegraphics[width=\textwidth, height=7cm]{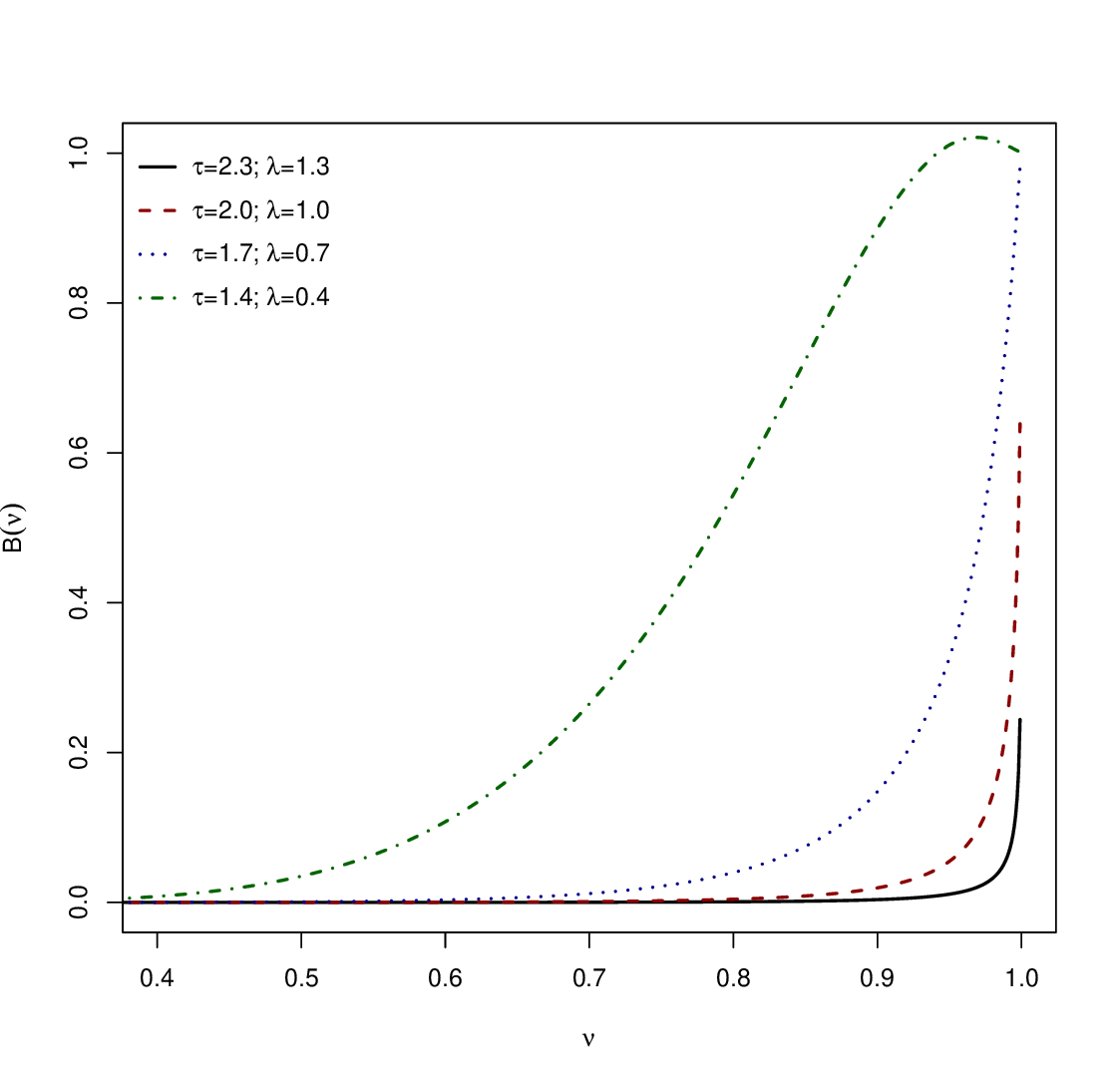}
			\end{minipage}
\hspace{.3cm}
			\begin{minipage}[c]{0.48\linewidth}
				\centering
				\includegraphics[width=\textwidth, height=7cm]{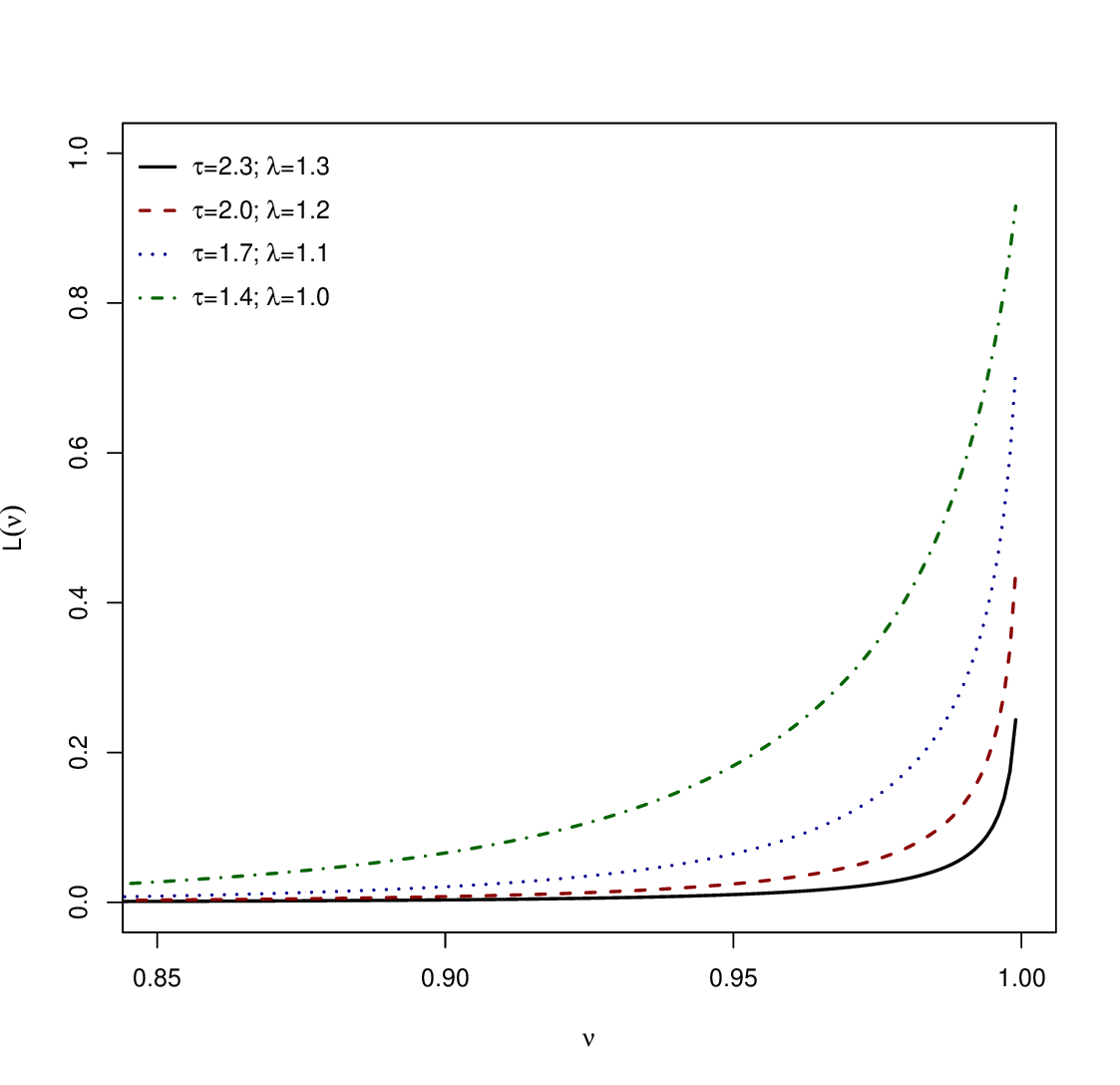}
			\end{minipage}
		\end{center}
		\caption{Boferroni and Lorenz curves of $\text{MOBXIIW}(\tau,\lambda,\alpha,\beta)$.}
	\label{Lorenz/BonferroniMOB}
	\end{figure}

	Incomplete moments find applications in the Bonferroni and Lorenz curves (for a probability $\nu$) as $B(\nu) = m_1(q)/\nu \mu^\prime_1$ and $L(\nu) = m_1(q)/\mu^\prime_1$, respectively, where $q = Q_X(\nu)$ is determined from (\ref{MOBXIIGqf}). Figure \ref{Lorenz/BonferroniMOB} illustrates these curves for the MOBXIIW distribution, with $\beta = 2.0$, $\alpha = 0.1$ and $\tau$ and $\lambda$ varying.

\subsection{Estimation}
	Let $x_1,\cdots,x_n$ be an independent and identically distributed (iid) random sample taken from pdf (\ref{MOBXIIpdf}). Then, the log-likelihood function for the parameter vector $\bm{\theta} = (\tau,\lambda,\bm{\xi})^\top$ reduces to 
\begin{align}\label{MOBXIImle}
\ell(\bm{\theta}) = &\,\,n\left[\tau\log(2) + \log(\tau) + \log(\lambda)\right] + \sum_{i = 1}^n \log\left[2 + G(x_i)^2\right] + (\tau - 1)\sum_{i = 1}^n \log G(x_i)\nonumber\\
&  \!- (\tau+1)\sum_{i = 1}^n\log\left\{2 - G(x_i)\left[1 + G(x_i)\right]\right\} - \sum_{i = 1}^n\log\left\{ 1\! + \!\left[\frac{2G(x_i)}{2 - G(x_i)\left[1 + G(x_i)\right]}\right]^\tau\right\}.
\end{align}
	
	Using current statistical programs, such as \texttt{R}, \texttt{Ox}, or \texttt{SAS}, it is possible to obtain the maximum likelihood estimate (MLE) of $\bm{\theta}$ by numerically maximizing Equation (\ref{MOBXIImle}). The AdequacyModel package \citep{Marinho2019} in \texttt{R} facilitates this process by offering a variety of maximization methods, such as Broyden-Fletcher-Goldfarb-Shannon (BFGS), Nelder-Mead, and Simulated Annealing (SANN).

\section{Regression model}\label{sec3MOBXII}
	A regression model for this family can be employed using the transformation $Y = \log(X)$, where $X$ has pdf (\ref{MOBXIIWpdf}). Thus, the density of $Y$ with $\beta = \text{e}^{\,\mu}$ and $\alpha = 1/\sigma$ has the form (for $y \in \text{I\!R}$)
\begin{align*}
f(y) =\,\,&\frac{ 2^{\tau}\,\tau\,\lambda\,\exp\left[\left(\frac{y-\mu}{\sigma}\right)-\text{e}^{\left(\frac{y-\mu}{\sigma}\right)}\right]\left\{2 + \left[1 - \text{e}^{-\text{e}^{\left(\frac{y-\mu}{\sigma}\right)}}\right]^2\right\}\left[1 - \text{e}^{-\text{e}^{\left(\frac{y-\mu}{\sigma}\right)}}\right]^{\tau-1}}{\sigma\left\{2 - \left[1 - \text{e}^{-\text{e}^{\left(\frac{y-\mu}{\sigma}\right)}}\right]\left[2 - \text{e}^{-\text{e}^{\left(\frac{y-\mu}{\sigma}\right)}}\right]\right\}^{\tau+1}}\nonumber\\
&\times\left\{1 + \left(\frac{2\left[1 - \text{e}^{-\text{e}^{\left(\frac{y-\mu}{\sigma}\right)}}\right]}{2 - \left[1 - \text{e}^{-\text{e}^{\left(\frac{y-\mu}{\sigma}\right)}}\right]\left[2 - \text{e}^{-\text{e}^{\left(\frac{y-\mu}{\sigma}\right)}}\right]}\right)^\tau\right\}^{-(\lambda+1)}\,,
\end{align*}
where $\sigma, \tau, \lambda > 0$ and $\mu \in \text{I\!R}$. The random variable $Z = (Y - \mu)/\sigma$ has density (for $z \in \text{I\!R}$)
\begin{align}\label{LMOBXIIzpdf}
f(z) = \,\,&\frac{ 2^{\tau}\,\tau\,\lambda\,\exp\left[z-\text{e}^{z}\right]\left\{2 + \left[1 - \text{e}^{-\text{e}^{z}}\right]^2\right\}\left[1 - \text{e}^{-\text{e}^{z}}\right]^{\tau-1}}{\left\{2 - \left[1 - \text{e}^{-\text{e}^{z}}\right]\left[2 - \text{e}^{-\text{e}^{z}}\right]\right\}^{\tau+1}}\nonumber\\
&\times\left\{1 + \left(\frac{2\left[1 - \text{e}^{-\text{e}^{z}}\right]}{2 - \left[1 - \text{e}^{-\text{e}^{z}}\right]\left[2 - \text{e}^{-\text{e}^{z}}\right]}\right)^\tau\right\}^{-(\lambda+1)}\,.
\end{align}

	Equation (\ref{LMOBXIIzpdf}) represents the standard LMOBXIIW distribution. A regression model for the response variable $y_i$ linked to a vector of explanatory variables $\bm{v}_i^\top = (v_{i1},\cdots,v_{ip})^\top$ can be expressed as
\begin{align}\label{regressionmodelLMOBXIIW}
y_i = \bm{v}_i^\top \bm{\eta} + \sigma z_i\,, \quad i = 1,\ldots,n\,,
\end{align}
where $\mu_i =\bm{v}_i^\top \bm{\eta}$, $\bm{\eta} = (\eta_1,\cdots,\eta_p)^\top$ is a vector of coefficients, and $z_i$ follows the density (\ref{LMOBXIIzpdf}).
	
	The log-likelihood function for $\bm{\zeta} = (\tau,\lambda,\sigma, \bm{\eta}^\top)^\top$ follows from Equations (\ref{LMOBXIIzpdf}) and (\ref{regressionmodelLMOBXIIW}) by considering $y_i=\min(Y_i, C_i)$, where $Y_i$ and $C_i$ are (assuming independence) the lifetime and non-informative censoring time, respectively. For right-censored data, it is given by 
\begin{align}\label{LMOBXIImle}
\ell(\bm{\zeta}) =&\,\,  d\left[-\log(\sigma) + \tau\log(2) + \log(\tau) + \log(\lambda)\right] + \sum^n_{i \in F} z_i - \sum^n_{i \in F}\text{e}^{z_i}\nonumber\\
& + \sum^n_{i \in F}\log\left\{2 + \left[1 - \text{e}^{-\text{e}^{z_i}}\right]^2\right\} + (\tau-1)\sum^n_{i \in F}\log\left[1 - \text{e}^{-\text{e}^{z_i}}\right]\nonumber\\
& - (\tau+1)\sum^n_{i \in F}\log\left\{2 - \left[1 - \text{e}^{-\text{e}^{z_i}}\right]\left[2 - \text{e}^{-\text{e}^{z_i}}\right]\right\}\nonumber\\
& -(\lambda+1) \sum^n_{i \in F}\log\left\{1 + \left(\frac{2\left[1 - \text{e}^{-\text{e}^{z_i}}\right]}{2 - \left[1 - \text{e}^{-\text{e}^{z_i}}\right]\left[2 - \text{e}^{-\text{e}^{z}}\right]}\right)^\tau\right\}\nonumber\\
& - \lambda \sum^n_{i \in C}\log\left\{1 + \left(\frac{2\left[1 - \text{e}^{-\text{e}^{z_i}}\right]}{2 - \left[1 - \text{e}^{-\text{e}^{z_i}}\right]\left[2 - \text{e}^{-\text{e}^{z_i}}\right]}\right)^\tau\right\}\,,
\end{align}
where $d$ is the number of failures, $z_i = (y_i-\mu_i)/\sigma$, and $F$ and $C$ denote the sets of lifetimes and censoring times, respectively. The MLE of $\bm{\zeta}$ can be found by numerically maximizing (\ref{LMOBXIImle})\,. Several numerical methods can be used for this task, such as BFGS, Nelder-Mead, and SANN.

\section{Simulations}\label{sec4MOBXII}
	
	The MOBXIIW distribution evaluates the MLEs of the new family. For this, random samples of four different sizes ($n = 50$, 100, 200, and 400) are generated using Equation (\ref{MOBXIIGqf}), considering three different parameter configurations. One thousand Monte Carlo replications calculate the average estimates (AEs), biases, and mean squared errors (MSEs). The Nelder-Mead numerical method maximizes the log-likelihood function for $\bm{\theta} = (\tau, \lambda, \beta, \alpha)^\top$ from the pdf (\ref{MOBXIIWpdf}) using the \texttt{optim} function in \texttt{R}.

\begin{table}[ht!]
\centering
\small
\caption{Simulations from the MOBXIIW distribution.}
\begin{tabular}{lccccccccccc}
\\[-1.8ex] \toprule
& & \multicolumn{3}{c}{(0.8, 2.5, 0.5, 3.5)} & \multicolumn{3}{c}{(2.8, 0.5, 1.5, 2.5)} & \multicolumn{3}{c}{(1.8, 0.3, 0.5, 0.9)}\\
\cmidrule(r){3-5}\cmidrule(r){6-8}\cmidrule(r){9-11}
$n$ & $\bm{\theta}$ & AE & Bias & MSE & AE & Bias & MSE & AE & Bias & MSE \\
\midrule
50  & $\tau$    & 0.6751 & -0.1249 & 0.4047  & 3.3180 & 0.5180  & 3.1074  & 2.3138 & 0.5138 & 1.5859\\
    & $\lambda$ & 2.1515 & -0.3485 & 5.4251  & 0.4880 & -0.0120 & 0.3777  & 0.3888 & 0.0888 & 0.2206\\
    & $\beta$   & 0.4549 & -0.0451 & 0.0151  & 1.4511 & -0.0489 & 0.0523  & 0.6250 & 0.1250 & 0.3222\\
    & $\alpha$  & 5.8388 & 2.3388 & 17.8879  & 3.0469 & 0.5469  & 2.1153  & 0.9374 & 0.0374 & 0.1201\\
        \\ 
100 & $\tau$    & 0.6762 & -0.1238 & 0.1931  & 3.0590 & 0.2590  & 1.4166  & 2.1441 & 0.3441 & 0.7857\\
    & $\lambda$ & 2.2820 & -0.2180 & 1.6331  & 0.4493 & -0.0507 & 0.1022  & 0.3752 & 0.0752 & 0.1560\\
    & $\beta$   & 0.4692 & -0.0308 & 0.0103  & 1.4516 & -0.0484 & 0.0348  & 0.6035 & 0.1035 & 0.2535\\
    & $\alpha$  & 5.1078 & 1.6078  & 8.7753  & 2.8691 & 0.3691  & 0.9872  & 0.9031 & 0.0031 & 0.0534\\
        \\
200 & $\tau$    & 0.7049 & -0.0951 & 0.0902  & 2.9508 & 0.1508  & 0.9012  & 1.9872 & 0.1872 & 0.2522\\
    & $\lambda$ & 2.4707 & -0.0293 & 2.9956  & 0.4582 & -0.0418 & 0.0558  & 0.3512 & 0.0512 & 0.0781\\
    & $\beta$   & 0.4822 & -0.0178 & 0.0094  & 1.4676 & -0.0324 & 0.0230  & 0.5644 & 0.0644 & 0.1207\\
    & $\alpha$  & 4.5611 & 1.0611  & 4.9997  & 2.7329 & 0.2329  & 0.6310  & 0.8889 &-0.0111 & 0.0285\\
       \\
400 & $\tau$    & 0.7293 & -0.0707 & 0.0657  & 2.9182 & 0.1182  & 0.5712  & 1.9002 & 0.1002 & 0.1073\\
    & $\lambda$ & 2.5194 & 0.0194  & 1.1950  & 0.4851 & -0.0149 & 0.0340  & 0.3362 & 0.0362 & 0.0443\\
    & $\beta$   & 0.4913 & -0.0087 & 0.0055  & 1.4885 & -0.0115 & 0.0140  & 0.5443 & 0.0443 & 0.0667\\
    & $\alpha$  & 4.2189 & 0.7189  & 2.5708  & 2.6043 & 0.1043  & 0.3861  & 0.8903 &-0.0097 & 0.0171\\
\bottomrule\label{TB1MOBXIIW}
\end{tabular}
\end{table}

	The results in Table \ref{TB1MOBXIIW} show that as the sample size increases, the biases and MSEs tend to decrease, and the AEs converge to the chosen parameter values in all scenarios. This indicates that the estimators of the new family are consistent. 

	To assess the MLEs of the regression model for the new family, a simulation study involving one thousand Monte Carlo replications is conducted. Sample sizes of $n = 50$, 100, 200, and 400 are generated from (\ref{MOBXIIGqf}), with $\mu_i = \eta_0 + \eta_1 v_{i1}$, where $v_{i1}$ follows a uniform distribution (0,1). The parameter values are: $\tau = 1.8$, $\lambda = 0.5$, $\sigma = 0.9$, $\eta_0 = 1.5$, and $\eta_1 = 2.2$. The censoring times $c_1,\cdots,c_n$ are generated from a uniform distribution (0,$b$), where $b$ determines the censoring percentage (0\%, 10\%, 30\%). The numerical optimization method Nelder-Mead maximizes Equation (\ref{LMOBXIImle}) using the \texttt{optim} function in \texttt{R}.

	The simulation process is described as (for $i = 1,\ldots,n$):
\begin{enumerate}
\item Generate $v_{i1} \sim \text{Uniform}\,(0,1)$ and set $\mu_i = \eta_0 + \eta_1 v_{i1}$.
\item Generate $y_i$ from (\ref{MOBXIIGqf}).
\item Generate $c_i \sim \text{Uniform}\,(0,b)$.
\item The observed times are $y^*_i = \min(y_i,c_i)$,  
where the censoring indicator $\delta_i = 1$ if $y_i \leq c_i$ 
and $\delta_i = 0$, otherwise. 
\end{enumerate}

\begin{table}[ht!]
\centering
\scalefont{0.9}
\caption{Simulations from the LMOBXIIW regression model.}
\begin{tabular}{lccccccccccc}
\\[-1.8ex]\toprule
& & \multicolumn{3}{c}{$0\%$} & \multicolumn{3}{c}{$10\%$} & \multicolumn{3}{c}{$30\%$}\\
\cmidrule(r){3-5}\cmidrule(r){6-8}\cmidrule(r){9-11}
$n$ & $\bm{\zeta}$ & AE & Bias & MSE & AE & Bias & MSE & AE & Bias & MSE \\
\midrule
50  & $\tau$     & 2.3477 & 0.5477 & 1.8787 & 2.2731 & 0.4731 & 1.8985 & 2.2396 & 0.4396 & 2.3321\\
    & $\lambda$  & 0.6052 & 0.1052 & 0.2160 & 0.6012 & 0.1012 & 0.3045 & 0.6591 & 0.1591 & 0.3103\\
    & $\sigma$   & 1.0198 & 0.1198 & 0.2939 & 0.9943 & 0.0943 & 0.2972 & 0.9604 & 0.0604 & 0.3668\\
    & $\eta_0$   & 1.5433 & 0.0433 & 0.1977 & 1.5298 & 0.0298 & 0.1897 & 1.5455 & 0.0455 & 0.2261\\
    & $\eta_1$   & 2.1986 &-0.0014 & 0.1130 & 2.1948 &-0.0052 & 0.1213 & 2.2011 & 0.0011 & 0.1856\\
\\
100 & $\tau$     & 2.1435 & 0.3435 & 0.8319 & 2.1930 & 0.3930 & 0.9577 & 2.1738 & 0.3738 & 1.0432\\
    & $\lambda$  & 0.5954 & 0.0954 & 0.1677 & 0.5724 & 0.0724 & 0.1632 & 0.6041 & 0.1041 & 0.2444\\
    & $\sigma$   & 1.0052 & 0.1052 & 0.2034 & 1.0085 & 0.1085 & 0.2020 & 0.9934 & 0.0934 & 0.2166\\
    & $\eta_0$   & 1.5394 & 0.0394 & 0.1451 & 1.5133 & 0.0133 & 0.1745 & 1.5140 & 0.0140 & 0.1745\\
    & $\eta_1$   & 2.2068 & 0.0068 & 0.0559 & 2.2024 & 0.0024 & 0.0611 & 2.2123 & 0.0123 & 0.0771\\
\\
200 & $\tau$     & 2.0372 & 0.2372 & 0.4214 & 2.0484 & 0.2484 & 0.4443 & 2.0916 & 0.2916 & 0.6137\\
    & $\lambda$  & 0.5720 & 0.0720 & 0.1230 & 0.5912 & 0.0912 & 0.1527 & 0.5927 & 0.0927 & 0.1364\\
    & $\sigma$   & 0.9951 & 0.0951 & 0.1290 & 0.9979 & 0.0979 & 0.1234 & 1.0170 & 0.1170 & 0.1671\\
    & $\eta_0$   & 1.5325 & 0.0325 & 0.1071 & 1.5461 & 0.0461 & 0.1163 & 1.5465 & 0.0465 & 0.1280\\
    & $\eta_1$   & 2.2014 & 0.0014 & 0.0267 & 2.2012 & 0.0012 & 0.0315 & 2.1997 &-0.0003 & 0.0409\\
\\
400 & $\tau$     & 1.9646 & 0.1646 & 0.2046 & 2.0037 & 0.2037 & 0.3043 & 2.0074 & 0.2074 & 0.3174\\
    & $\lambda$  & 0.5562 & 0.0562 & 0.0771 & 0.5598 & 0.0598 & 0.0840 & 0.5588 & 0.0588 & 0.0932\\
    & $\sigma$   & 0.9712 & 0.0712 & 0.0617 & 0.9884 & 0.0884 & 0.0910 & 0.9858 & 0.0858 & 0.0921\\
    & $\eta_0$   & 1.5271 & 0.0271 & 0.0807 & 1.5271 & 0.0271 & 0.0886 & 1.5267 & 0.0267 & 0.0990\\
    & $\eta_1$   & 2.1945 &-0.0055 & 0.0129 & 2.1960 &-0.0040 & 0.0137 & 2.1889 &-0.0111 & 0.0183\\
\bottomrule\label{TB2LMOBXIIW}
\end{tabular}
\end{table}

The results in Table \ref{TB2LMOBXIIW} indicate that the MLEs of the regression model are consistent, with the AEs converging to the true parameter values. As the sample size increases, both the biases and MSEs decrease. While increasing the censoring percentage impacts some parameters more, this effect diminishes in large samples.

\section{Applications}\label{sec5MOBXII}
	This section shows the versatility and potential of the proposed family by applying the MOBXIIW distribution and its associated regression model to three data sets. The quality of the fit provided by each model is assessed using the Cramér-von Mises $(W^*)$ and Anderson-Darling $(A^*)$ statistics \citep{chen1995}, as well as the Akaike Information Criterion (AIC), the Consistent AIC (CAIC), the Bayesian IC (BIC), the Hannan-Quinn IC (HQIC), and the Kolmogorov-Smirnov (KS) statistic (along with its corresponding $p$-value). The lower value for these measures indicates a better fit to the data.

\subsection{Dengue data}\label{sec6.1}
	The data consists of 345 observations on the number of confirmed dengue cases in April 2024 within São Paulo State, Brazil. This data can be extracted from the link \url{https://saude.sp.gov.br/cve-centro-de-vigilancia-epidemiologica-prof.-alexandre-vranjac/oldzoonoses/dengue/dados-estatisticos}, and reveals an average number of confirmed cases of 76.884, with a standard deviation of 75.793. Furthermore, the skewness of 0.903 and the kurtosis of 2.681 indicate that the data are right-skewed and platykurtic. According to \citep{de2024vacina}, dengue is a disease caused by the dengue virus, transmitted mainly by the Aedes aegypti mosquito. Although many cases are asymptomatic, the disease can cause fever, body aches, skin spots, and other complications. Some patients recover without intervention, but severe cases, such as dengue hemorrhagic, necessitate medical attention and can potentially lead to death. In Brazil, records of dengue date back to the 19th century, with initial epidemics occurring in São Paulo and Rio de Janeiro.
	
\begin{table}[!ht] 
\centering 
\caption{MLEs and SEs of the fitted models to dengue data.} 
\label{TBdengue} 
\begin{tabular}{@{\extracolsep{5pt}} lccccc} 
\\[-1.8ex] \toprule
\multicolumn{1}{c}{Distribution} & \multicolumn{4}{c}{MLEs (SEs)}\\
\midrule
MOBXIIW$(\tau,\lambda,\beta,\alpha)$& 0.252   & 1.935   & 131.974 & 3.182   \\ 
                                    & (0.044) & (0.211) & (12.176)& (0.553) \\
\midrule 
KW$(a,b,\beta,\alpha)$              & 1.056   & 0.095   & 0.830   & 4.117   \\ 
                                    & (0.052) & (0.006) & (0.012) & (0.008) \\
\midrule 
BW$(a,b,\beta,\alpha)$              & 0.798   & 0.090   & 0.891   & 5.063   \\ 
                                    & (0.096) & (0.005) & (0.002) & (0.005) \\
\midrule 
WW$(\tau,\lambda,\beta,\alpha)$     & 0.012   & 0.679   & 0.210   & 0.011   \\ 
                                    & (0.003) & (0.133) & (0.016) & (0.003) \\ 
\midrule  
LW$(\tau,\lambda,\beta,\alpha)$     & 0.089   & 0.594   & 0.842   & 4.221   \\ 
                                    & (0.005) & (0.150) & (0.006) & (0.006) \\
\midrule    
WE$(\beta,\alpha)$                  & 0.014   & 0.832                       \\ 
                                    & (0.001) & (0.037)                     \\ 
\bottomrule
\end{tabular} 
\end{table} 

	The MOBXIIW distribution is then compared with other established distributions, including the Kumaraswamy Weibull (KW) \citep{cordeiro2010kumaraswamy}, beta Weibull (BW) \citep{Famoye2005}, Weibull Weibull (WW) \citep{Bourguignon2014},  Lomax Weibull (LW) \citep{cordeiro2019odd}, and Weibull (WE). Thus, Table \ref{TBdengue} presents the MLEs and standard errors (SEs) for the distributions applied to dengue data, all providing precise estimates. The MOBXIIW distribution demonstrates superior performance, as indicated by lower adequacy measure values in Table \ref{TBdengueAD}.

\begin{figure}[!ht]
		\begin{center}
			\begin{minipage}[c]{0.48\linewidth}
				\centering
				\includegraphics[width=\textwidth, height=7cm]{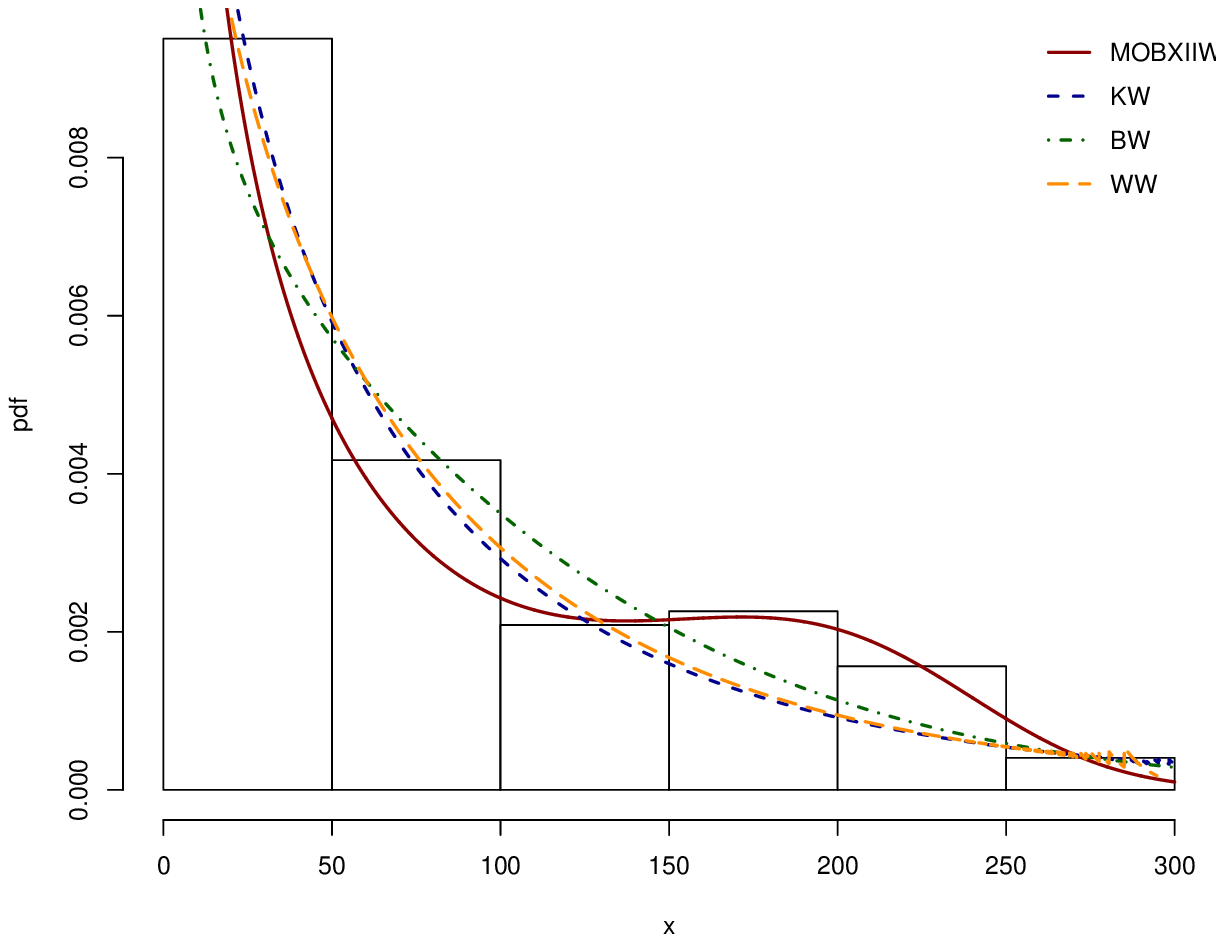}
			\end{minipage}
\hspace{.3cm}
			\begin{minipage}[c]{0.48\linewidth}
				\centering
				\includegraphics[width=\textwidth, height=7cm]{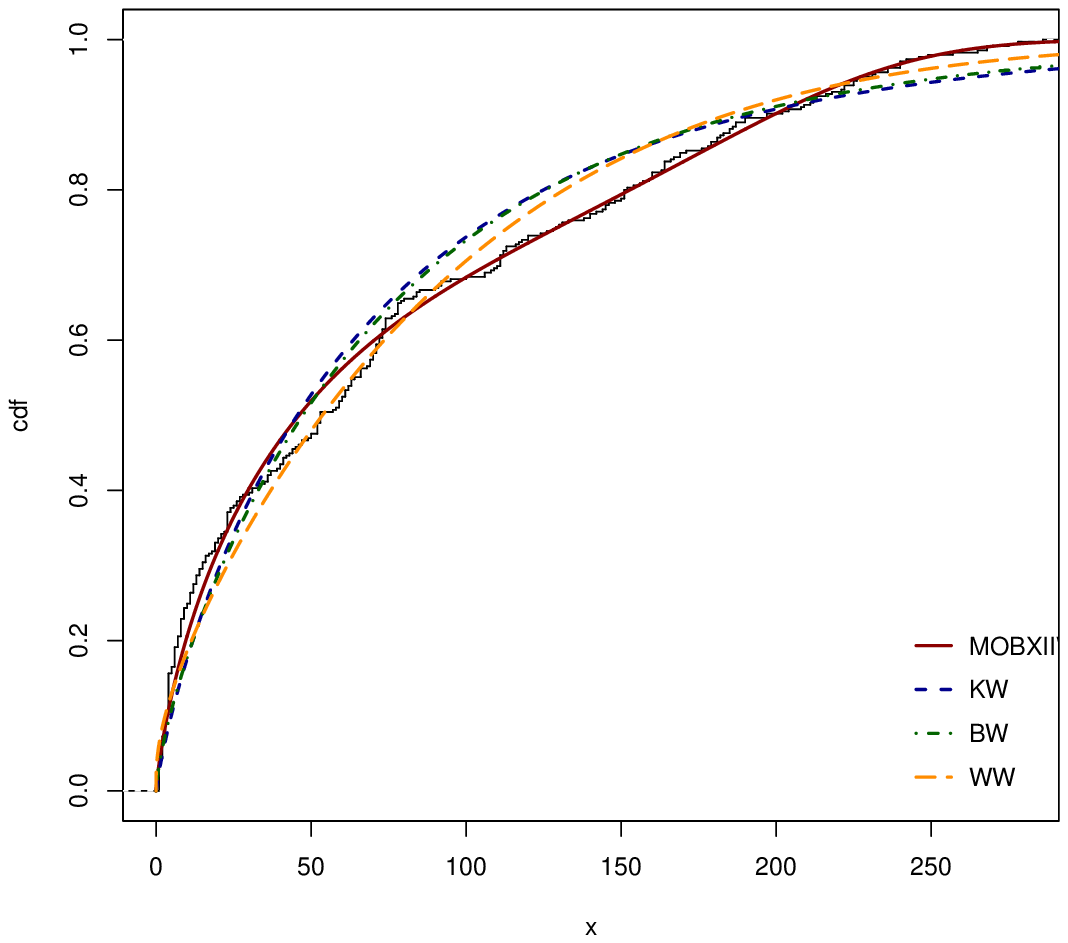}
			\end{minipage}
		\end{center}
		\caption{Some estimated pdfs and cdfs for dengue data.}
	\label{denguedata}
	\end{figure}

	The generalized likelihood ratio (GLR) test \citep{Vuong1989} compares the MOBXIIW model with the KW (GLR = 14.444), BW (GLR = 14.284), WW (GLR = 17.486), LW (GLR = 15.119), and WE (GLR = 14.531) models at a significance level of $5\%$. The GLR test results indicate that the MOBXIIW model provides a superior fit to the data compared to the alternatives. Figure \ref{denguedata} visually confirms this through the close correspondence between the pdf and cdf estimated by the model and the histogram and empirical cdf of the data.

	All previous results are obtained using the \texttt{AdequacyModel} package in \texttt{R}, with the numerical method BFGS.

\begin{table}[!ht] 
\centering 
\caption{Adequacy measures of the fitted models to dengue data.} 
\label{TBdengueAD} 
\scalefont{0.87}
\begin{tabular}{@{\extracolsep{5pt}} lcccccccc} 
\\[-1.8ex] \toprule
Distribution & $W^*$ & $A^*$ & AIC & CAIC & BIC & HQIC & KS & $p$-value \\ 
\midrule 
MOBXIIW & 0.203 & 1.370 & 3621.134 & 3621.252 & 3636.508 & 3627.257 & 0.052 & 0.290 \\ 
KW      & 0.618 & 4.172 & 3675.727 & 3675.845 & 3691.101 & 3681.850 & 0.081 & 0.021 \\ 
BW      & 0.569 & 3.961 & 3674.042 & 3674.160 & 3689.417 & 3680.165 & 0.079 & 0.028 \\ 
WW      & 0.544 & 3.988 & 3681.351 & 3681.469 & 3696.726 & 3687.474 & 0.071 & 0.059 \\ 
LW      & 0.559 & 3.841 & 3670.617 & 3670.734 & 3685.991 & 3676.740 & 0.069 & 0.070 \\ 
WE      & 0.613 & 4.155 & 3671.720 & 3671.755 & 3679.407 & 3674.781 & 0.078 & 0.030 \\ 
\bottomrule
\end{tabular} 
\end{table} 

\subsection{Length of stay data in Japan}
	A bimodal data set is employed to evaluate the flexibility of the new family. This data set comprises the length of stay in years for 147 Brazilian immigrants in Japan in 2010 \citep{bortolini2017extended}. The average length of stay in Japan is 12.81 years and a standard deviation of 6.146. The data are left-skewed and platykurtic according to the skewness (-0.3552) and kurtosis (1.7899) values.
	
	In this study, the comparative analysis extends beyond the previously described distributions. It incorporates three additional distributions known for their ability to handle bimodal data: the Kumaraswamy flexible Weibull (KFW) \citep{Sobh2016}, the Marshall-Olkin Weibull (WMOW) \citep{korkmaz2019weibull}, and the extended Weibull log-logistic (EWLL) \citep{abouelmagd2019extended}. This analysis clarifies how the flexibility of the MOBXIIW distribution in fitting bimodal data compares to alternatives in the literature.
	
\begin{figure}[!ht]
		\begin{center}
			\begin{minipage}[c]{0.48\linewidth}
				\centering
				\includegraphics[width=\textwidth, height=7cm]{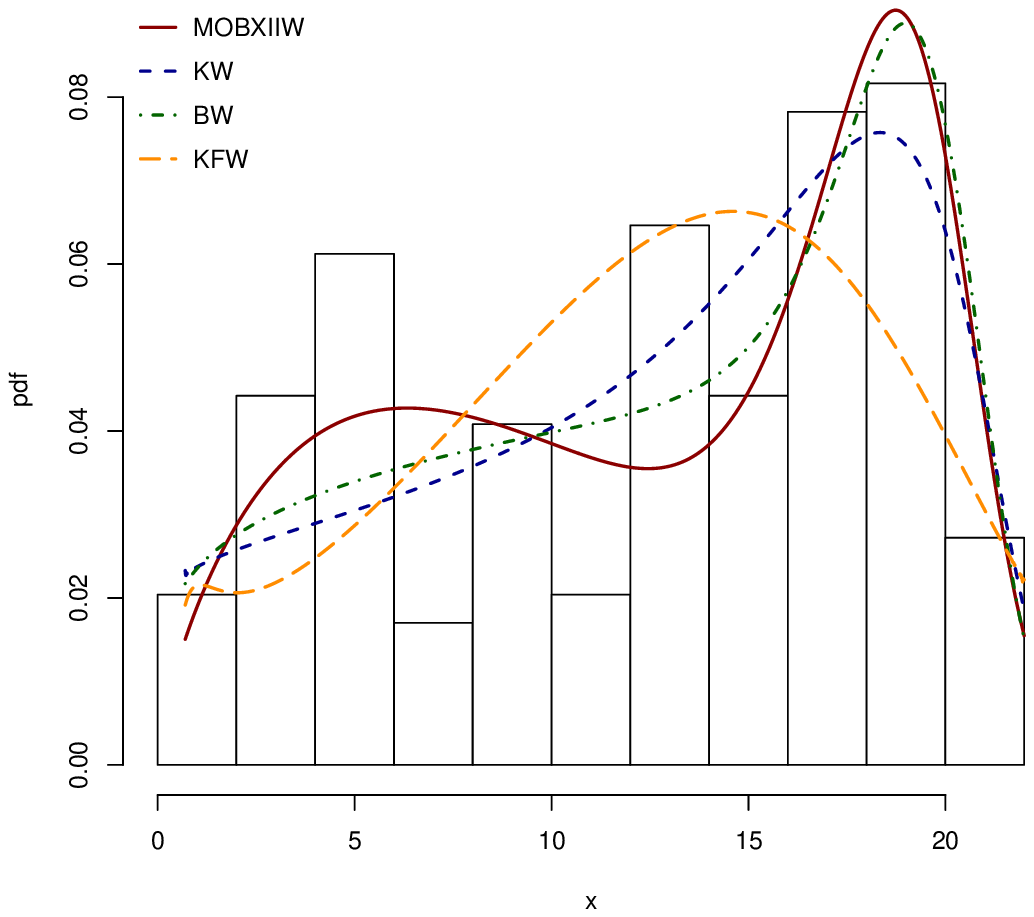}
			\end{minipage}
\hspace{.3cm}
			\begin{minipage}[c]{0.48\linewidth}
				\centering
				\includegraphics[width=\textwidth, height=7cm]{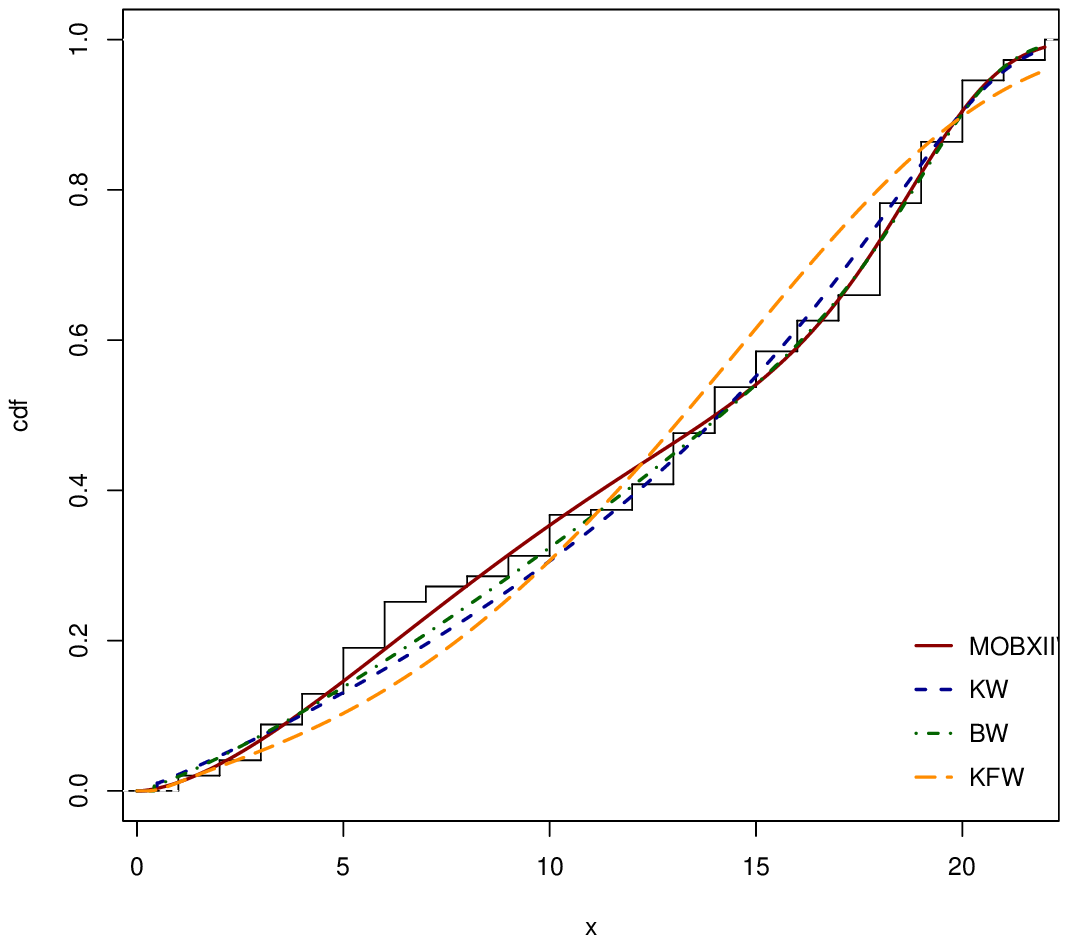}
			\end{minipage}
		\end{center}
		\caption{Some estimated pdfs, and cdfs for the data on length of stay in Japan.}
	\label{japandata}
	\end{figure}

	Table \ref{TBjapan} provides the MLEs and SEs for the selected models. Except for the LW and WMOW distributions, all the others yield accurate estimates. Notably, the MOBXIIW distribution exhibits the lowest adequacy measure values, as illustrated in Table \ref{TBjapanAD}. The GLR tests comparing the MOBXIIW model with the KFW (GLR = 6.779), KW (GLR = 5.333), BW (GLR = 4.437), WW (GLR = 8.077), LW (GLR = 6.899), WMOW (GLR = 9.409), and EWLL (GLR = 8.370) models at a significance level of $5\%$ confirm the superior fit of the MOBXIIW model to the data. Figure \ref{japandata} shows that the estimated pdf and cdf of the MOBXIIW distribution closely match the histogram and empirical cdf of the data compared to the alternatives. Again, all the results in this subsection are derived using the \texttt{AdequacyModel} package in \texttt{R}, with the numerical method BFGS.

\begin{table}[!ht] 
\centering 
\caption{MLEs and SEs of the models fitted to the data on length of stay in Japan.} 
\label{TBjapan} 
\begin{tabular}{@{\extracolsep{5pt}} lccccc} 
\\[-1.8ex] \toprule
\multicolumn{1}{c}{Distribution} & \multicolumn{4}{c}{MLEs (SEs)}\\
\midrule
MOBXIIW$(\tau,\lambda,\beta,\alpha)$ & 0.230  & 1.022  & 14.582 & 7.275  \\ 
                                     & (0.043)  & (0.189)  & (0.921)  & (1.323)  \\ 
\midrule
KFW$(a,b,\beta,\alpha)$              & 2.512  & 0.099  & 0.161  & 0.791  \\ 
                                     & (0.113)  & (0.011)  & (0.004)  & (0.007)  \\ 
\midrule                                     
KW$(a,b,\beta,\alpha)$               & 0.105  & 0.498  & 10.315 & 18.313 \\ 
                                     & (0.019)  & (0.132)  & (0.036)  & (0.799)  \\ 
\midrule                                     
BW$(a,b,\beta,\alpha)$               & 0.120  & 0.154  & 10.247 & 15.997 \\ 
                                     & (0.012)  & (0.019)  & (0.239)  & (0.072)  \\ 
\midrule                                     
WW$(\tau,\lambda,\beta,\alpha)$      & 0.013  & 0.153  & 0.656  & 0.094  \\ 
                                     & (0.004)  & (0.010)  & (0.003)  & (0.002)  \\ 
\midrule                                     
LW$(\tau,\lambda,\beta,\alpha)$      & 19.776 & 71.849 & 1.456  & 11.359 \\ 
                                     & (16.758) & (59.208) & (0.202)  & (2.435)  \\ 
\midrule                                     
WMOW$(\alpha,\beta,\gamma,\theta)$   & 6.000  & 0.517  & 3.292  & 12.082 \\ 
                                     & (3.391)  & (0.158)  & (1.084)  & (1.379)  \\
\midrule                                      
EWLL$(\lambda,\alpha,\beta)$         & 0.286  & 0.606  & 8.784                   \\ 
                                     & (0.081)  & (0.078)  & (2.046)             \\ 
\bottomrule 
\end{tabular} 
\end{table} 

\begin{table}[!ht] 
\centering 
\caption{Adequacy measures of the models fitted to the data on length of stay in Japan.} 
\label{TBjapanAD} 
\scalefont{0.87}
\begin{tabular}{@{\extracolsep{5pt}} lcccccccc} 
\\[-1.8ex] \toprule
Distribution & $W^*$ & $A^*$ & AIC & CAIC & BIC & HQIC & KS & $p$-value \\ 
\midrule 
MOBXIIW & 0.093 & 0.597 & 899.939 & 900.221 & 911.901 & 904.799 & 0.072 & 0.421 \\ 
KFW     & 0.407 & 2.566 & 930.342 & 930.623 & 942.303 & 935.202 & 0.142 & 0.005 \\ 
KW      & 0.161 & 1.055 & 906.680 & 906.962 & 918.642 & 911.541 & 0.098 & 0.116 \\ 
BW      & 0.104 & 0.733 & 901.701 & 901.982 & 913.662 & 906.561 & 0.078 & 0.321 \\ 
WW      & 0.442 & 2.892 & 940.992 & 941.274 & 952.954 & 945.852 & 0.134 & 0.001 \\ 
LW      & 0.395 & 2.471 & 930.209 & 930.490 & 942.170 & 935.069 & 0.138 & 0.007 \\ 
WMOW    & 0.522 & 3.286 & 948.887 & 949.169 & 960.849 & 953.747 & 0.135 & 0.009 \\ 
EWLL    & 0.819 & 4.800 & 961.541 & 961.709 & 970.512 & 965.186 & 0.154 & 0.001  \\  
\bottomrule
\end{tabular} 
\end{table} 

\subsection{COVID-19 data}
	To evaluate the adequacy of the regression model for the new family, a data set is selected that pertains to the lifetimes of 956 individuals with COVID-19 in Fortaleza, the capital of Ceará, in 2023. This data is available at the following link: \url{https://opendatasus.saude.gov.br/en/dataset/notificacoes-de-sindrome-gripal-leve-2022}. The average lifetime of these individuals is 17.387 days, with a standard deviation of 8.930. The skewness of 0.414 and the kurtosis of 2.653 indicate that the data are right-skewed and platykurtic. Here, the response variable $y_i$ represents the survival time from symptom onset to death due to COVID-19 (failure).
	
	Approximately 58.78\% of the observations are censored, indicating that they pertain to individuals who either died from causes unrelated to COVID-19 or survived until the end of the study. The variables considered (for $i = 1, \ldots, 956$) include: $\delta_i\!:$ censoring indicator (0 = censored, 1 = observed lifetime), $v_{i1}\!:$ age (in years), and $v_{i2}\!:$ hepatic disease (1 = yes, 0 = no or not informed). 
	
\begin{figure}
		\begin{center}
			\begin{minipage}[c]{0.48\linewidth}
				\centering
				(a)
				\includegraphics[width=\textwidth, height=7cm]{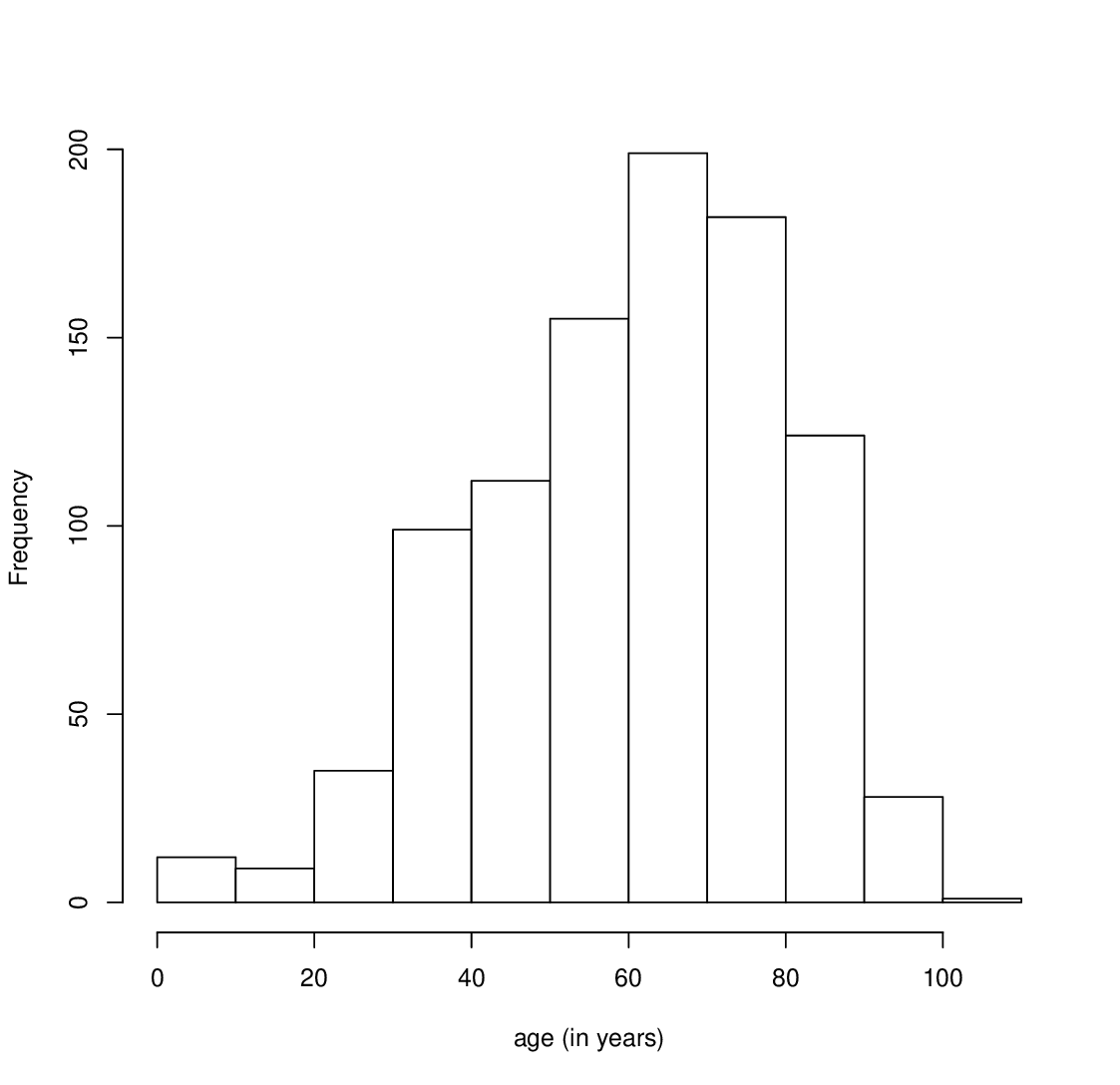}
			\end{minipage}
\hspace{.3cm}
			\begin{minipage}[c]{0.48\linewidth}
				\centering
				(b)
				\includegraphics[width=\textwidth, height=7cm]{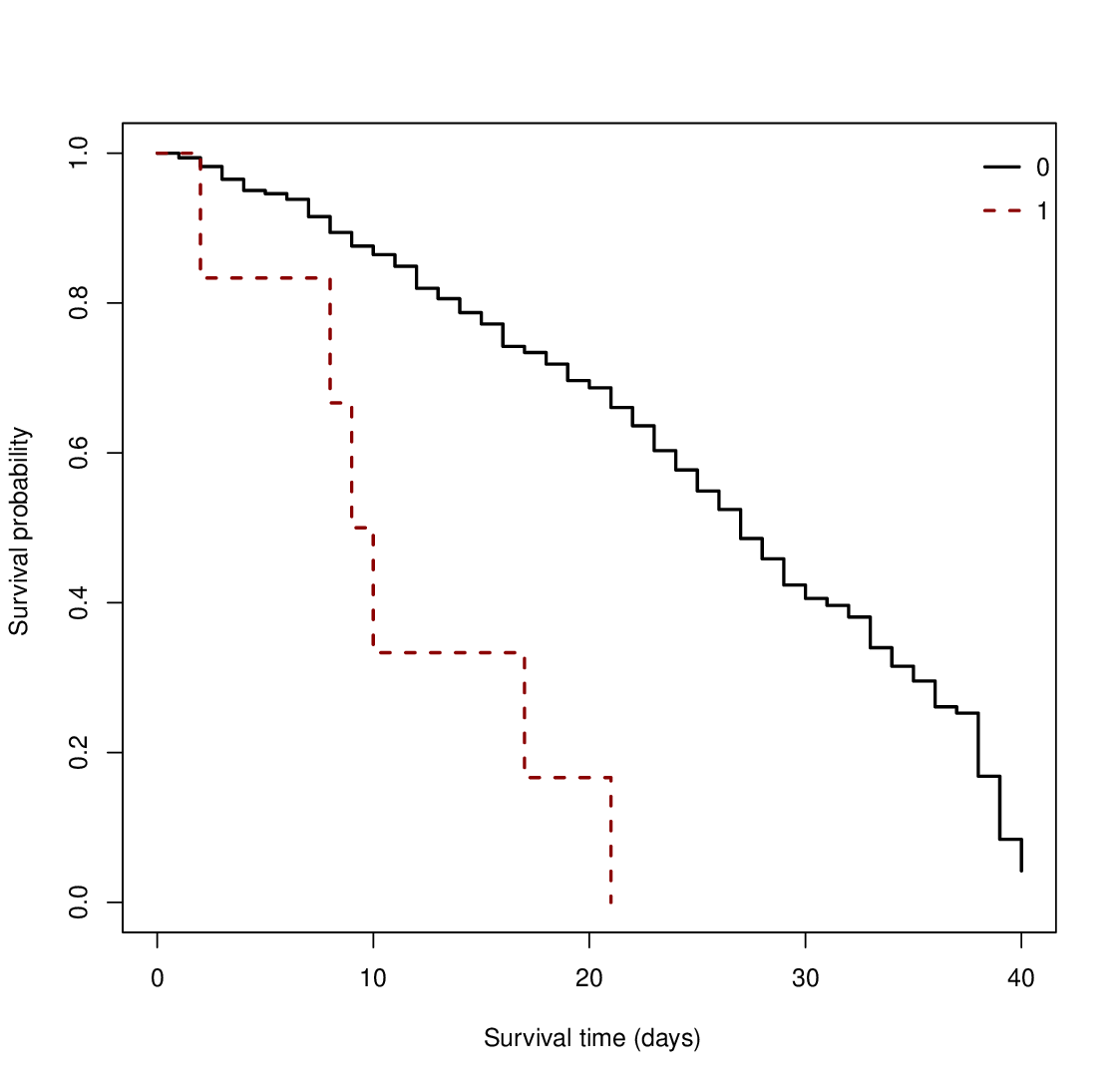}
			\end{minipage}
		\end{center}
		\caption{Histogram for age (a) and Kaplan-Meier curves for hepatic disease (b) for COVID-19 data in Fortaleza.}
\label{coviddata}
\end{figure}

	Figure \ref{coviddata}(a) shows that the majority of patients are in the 40-80 age group, with a notable peak in the 60-70 age range, indicating greater vulnerability in this group. Figure \ref{coviddata}(b) highlights the difference between patients with and without hepatic disease. The dashed curve (group 1) exhibits a steeper decline compared to the solid curve (group 0), suggesting a lower probability of survival for group 1.
	
\begin{figure}[!ht]
		\begin{center}
			\begin{minipage}[c]{0.48\linewidth}
				\centering
				\includegraphics[width=\textwidth, height=7cm]{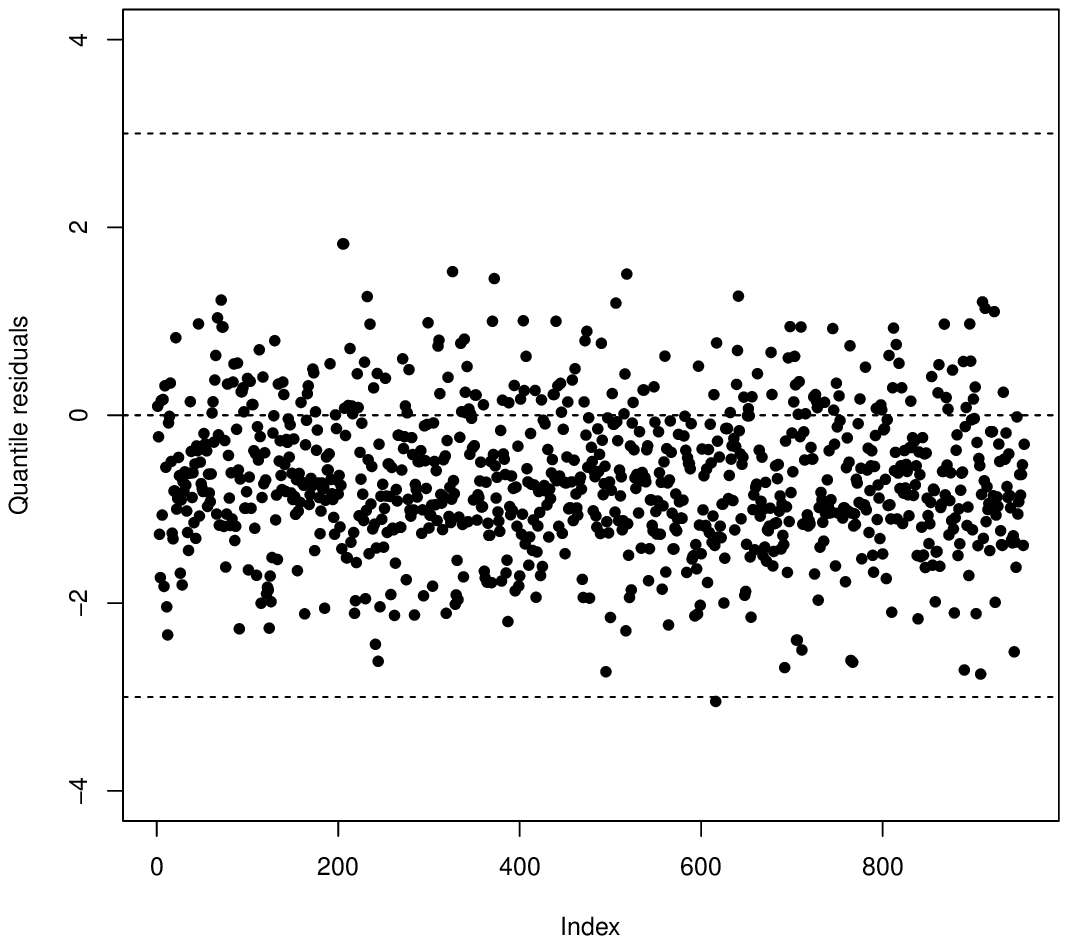}
			\end{minipage}
\hspace{.3cm}
			\begin{minipage}[c]{0.48\linewidth}
				\centering
				\includegraphics[width=\textwidth, height=7cm]{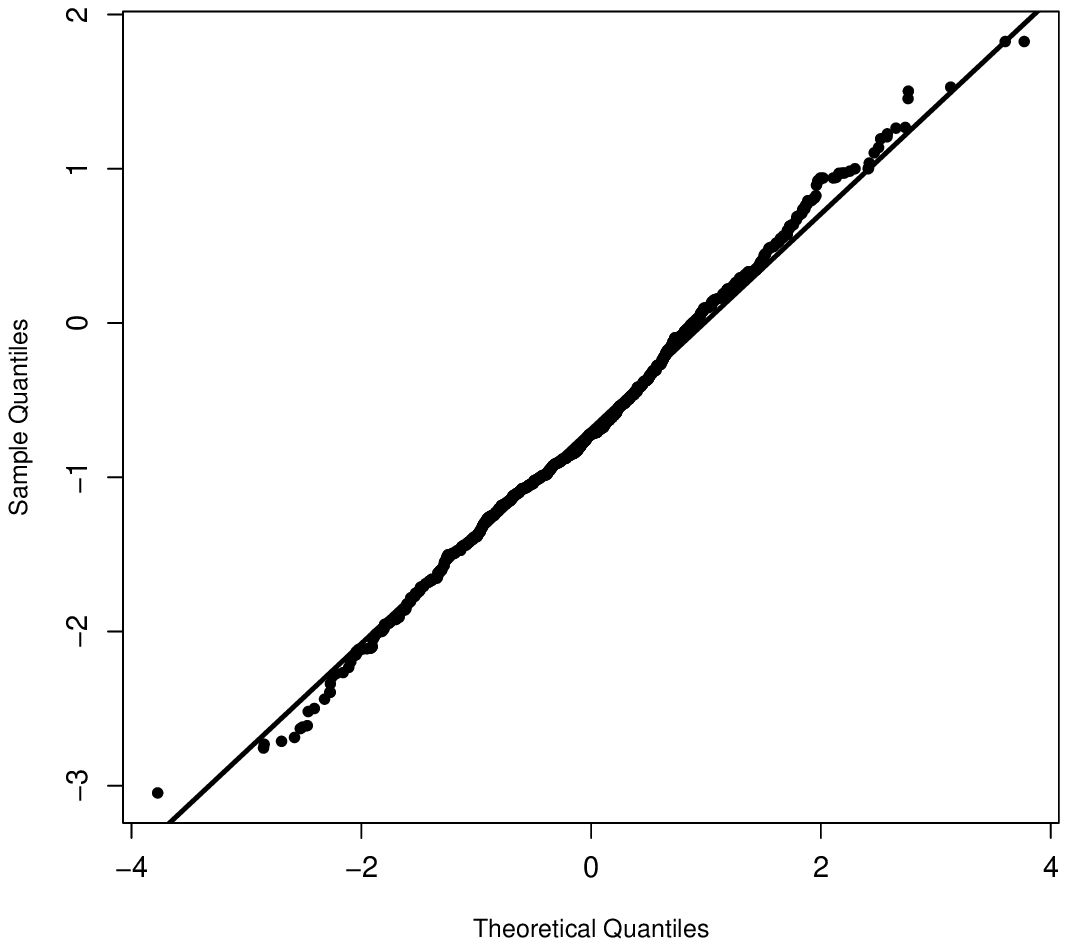}
			\end{minipage}
		\end{center}
		\caption{(a) Index plot and (b) normal probability plot of the qrs for COVID-19 data in Fortaleza.}
	\label{coviddataresiduals}
	\end{figure}

	Then, the regression model for these data is
\begin{align*}
y_i = \eta_0 + \eta_1v_{i1} + \eta_2v_{i2} + \sigma z_i, \,\, i = 1,\ldots,956\,,
\end{align*}
where $z_i$ has pdf (\ref{LMOBXIIzpdf}). The results are compared with three regression models: the log-Kumaraswamy Weibull (LKW), log-beta Weibull (LBW) \citep{ortega2013log}, and log-Weibull Weibull (LWW) regressions. The numbers in Table \ref{TB7COVID} show that the explanatory variables age and hepatic disease are significant at the 5\% level. Negative values of $\eta_1$ and $\eta_2$ indicate that increasing age or the presence of hepatic disease is associated with shorter failure times. Furthermore, the lowest values of the adequacy measures in Table \ref{TB8COVID} suggest that the LMOBXIIW regression provides a superior fit to the current data compared to the alternatives. To analyze the residuals of the new adjusted regression, quantile residuals (qrs) are employed, following \citep{dunn1996}.
\begin{align*}
qr_i = \Phi^{-1}\left(1 - \left\{1 + \left[\frac{2\left(1 - \text{e}^{-\text{e}^{\hat{z}_i}}\right)}{2 - \left(1 - \text{e}^{-\text{e}^{\hat{z}_i}}\right)\left(2 - \text{e}^{-\text{e}^{\hat{z}_i}}\right)}\right]^{\hat{\tau}}\right\}^{-\hat{\lambda}}\right)\,,
\end{align*}
where $\Phi^{-1}(\cdot)$ represents the inverse cdf of the standard normal distribution, \(\hat{z}_i = (y_i - \hat{\mu}_i)/\hat{\sigma}\), and \(\hat{\mu}_i = \bm{v}_i^\top \hat{\bm{\eta}} \). As shown in Figure \ref{coviddataresiduals}, the qrs display a random pattern and asymptotically align with the standard normal distribution, confirming that the LMOBXIIW regression model fits the data well. All results are calculated using an \texttt{R} script with the BFGS numerical optimization method in the \texttt{optim} function.

\begin{table}[!ht] 
\centering 
\caption{Estimates for COVID-19 data in Fortaleza.} 
\label{TB7COVID} 
\begin{tabular}{@{\extracolsep{5pt}} lcccccccccc} 
\\[-1.8ex] \toprule
 Model  & $\tau$ & $\lambda$ & $\sigma$ & $\eta_0$  & $\eta_1$ & $\eta_2$  \\ 
\midrule
         & 0.791   & 0.284   & 0.491  & 3.432   & -0.011 & -0.858 \\ 
LMOBXIIW & (0.102) & (0.144) & (0.063)& (0.250) & (0.002)  & (0.197)\\
         &         &         &        & $[<0.001]$&$[<0.001]$ & $[<0.001]$ \\
      \midrule 
         & 0.535   & 0.250   & 0.401  & 3.671 & -0.011 & -0.851 \\ 
LKW      &(0.149)  & (0.217) & (0.102)& (0.469) & (0.002)  & (0.188) \\
	     &         &         &        & $[<0.001]$ & $[<0.001]$ & $[<0.001]$\\
	    \midrule 
	     & 0.457   & 0.234   & 0.318  & 3.693   & -0.009 & -0.739   \\ 
LBW      & (0.310) & (0.224) & (0.211)& (0.505) & (0.004) & (0.192)  \\
	     &         &         &        & $[<0.001]$ &  [0.047] & $[<0.001]$ \\
	    \midrule
         & 0.099   &  1.933  & 1.731  &  3.488  & -0.010  &  -0.824 \\ 
LWW      & (0.175) & (1.254) & (1.298) & (0.688)  & (0.002) & (0.189) \\
	     &         &         &    & $[<0.001]$ & $[<0.001]$ & [0.018]\\
\bottomrule
\end{tabular} 
\end{table} 
	
\begin{table}[!ht] 
\centering 
\caption{Adequacy measures for COVID-19 data in Fortaleza.} 
\label{TB8COVID} 
\begin{tabular}{@{\extracolsep{5pt}} lcccc} 
\\[-1.8ex] \toprule 
Model & AIC & CAIC & BIC & HQIC\\ 
\midrule
LMOBXIIW & 1495.355 & 1495.507 & 1524.532 & 1506.468 \\ 
LKW      & 1497.556 & 1497.708 & 1526.732 & 1508.669 \\ 
LBW      & 1496.884 & 1497.036 & 1526.061 & 1507.998 \\ 
LWW      & 1499.091 & 1499.243 & 1528.268 & 1510.205 \\ 
\bottomrule 
\end{tabular} 
\end{table}

\section{Conclusions}\label{sec6MOBXII}
	The modified odd Bur XII-G (MOBXII-G) family, which extended the modeling capabilities of its baseline distributions by accommodating bimodal and bathtub-shaped shapes, was presented. A regression model for censored data was also built. Maximum likelihood estimators of the new models were found to be consistent through simulation. Evaluating the performance of MOBXII-G models using three real data sets revealed that, particularly for bimodal data sets, this new distribution outperformed well-known families such as Kumaraswamy-G and beta-G. 

\section*{Acknowledgments}
Fundação de Amparo à Ciência e Tecnologia do Estado 
de Pernambuco (FACEPE) [IBPG-1448-1.02/20] supports this work. 

\bibliographystyle{apalike}
\bibliography{referencias}
\end{document}